\documentclass[preprint, amsmath, amssymb, nobibnotes, aps, prb, superscriptaddress, citeautoscript]{revtex4-2}
\usepackage{import}
\usepackage{preamble}
\usepackage{soul}
\usepackage{float}
\usepackage[normalem]{ulem}
\usepackage[version=4]{mhchem}

\def\QE{\textsc{Quantum ESPRESSO}\,}

\newcommand{\editorr}[2]{%
  \expandafter\newcommand\csname #1note\endcsname[1]{%
    \textcolor{#2}{(\textbf{#1:} ##1)}}%
  \expandafter\newcommand\csname #1\endcsname[1]{%
    \textcolor{#2}{##1}}%
  \expandafter\newcommand\csname #1cancel\endcsname[1]{%
    \textcolor{#2}{\sout{##1}}}%
  \expandafter\newcommand\csname #1change\endcsname[2]{%
    \textcolor{#2}{\sout{##1} ##2}}%
  \newenvironment{#1text}{\color{#2}}{\color{black}}
}

\begin{document}

\title{Teaching oxidation states to neural networks}

\author{Cristiano Malica}
\email{cmalica@uni-bremen.de}
\affiliation{U Bremen Excellence Chair, Bremen Center for Computational Materials Science, and MAPEX Center for Materials and Processes, University of Bremen, D-28359 Bremen, Germany}
% ...
\author{Nicola Marzari}
\affiliation{U Bremen Excellence Chair, Bremen Center for Computational Materials Science, and MAPEX Center for Materials and Processes, University of Bremen, D-28359 Bremen, Germany}

\affiliation{Theory and Simulation of Materials (THEOS), and National Centre for Computational Design and Discovery of Novel Materials (MARVEL), \'Ecole Polytechnique F\'ed\'erale de Lausanne (EPFL), CH-1015 Lausanne, Switzerland}

\affiliation{Laboratory for Materials Simulations, Paul Scherrer Institut, 5232 Villigen PSI, Switzerland}

\begin{abstract}

{\color{black}
While the accurate description of redox reactions remains a challenge for first-principles calculations, it has been shown that extended Hubbard functionals (DFT+U+V) can provide a reliable approach, mitigating self-interaction errors, in materials with strongly localized $d$ or $f$ electrons. Here, we first show that DFT+U+V molecular dynamics is capable to follow the adiabatic evolution of oxidation states over time, using representative Li-ion cathode materials.  In turn, this allows to develop redox-aware machine-learned potentials. We show that considering atoms with different oxidation states (as accurately predicted by DFT+U+V) as distinct species in the training leads to potentials that are able to identify the correct ground state and pattern of oxidation states for redox elements present. This can be achieved, e.g., through a systematic combinatorial search for the lowest energy configuration or with stochastic methods. This brings the advantages of machine-learned potentials to key technological applications (e.g., rechargeable batteries), which require an accurate description of the evolution of redox states.   
}

\end{abstract}

\flushbottom
\date{\today}
\maketitle
\thispagestyle{empty}

\section*{INTRODUCTION}
The concept of oxidation state is ubiquitous in physics and chemistry. The evolution of oxidation states on the atoms of a system follows redox reactions, electrolysis, and many other crucial electrochemical processes. These processes are essential for understanding numerous phenomena underlying contemporary technologies {\color{black} ~\cite{Walsh:2018, Barton:2020, Jablonka:2021}. } Redox reactions, for example, are at the heart of the functioning of rechargeable batteries, enabling the conversion and storage of energy {\color{black} ~\cite{Nykvist:2015, Sarma:2018, Ponrouch:2019, Liu:2019} }. Modeling these processes is fundamental for understanding the underlying physical mechanism, allowing the identification and optimization of key factors influencing the behaviour of materials, and laying the groundwork for the development of possible innovative solutions. 

In this domain, density-functional theory (DFT) ~\cite{Hohenberg:1964,Kohn:1965} plays a fundamental role, enabling first-principles modeling of atomic systems by drawing on the fundamental tenets of quantum mechanics. This approach has led to significant advances in the analysis, characterization, and discovery of materials. 
Moreover, the combination of DFT with molecular dynamics {\color{black} (i.e., first-principles molecular dynamics, FPMD)} has further extended first-principles studies from static properties (at zero temperature) to dynamic properties (at finite temperature), providing a comprehensive and general investigative tool~\cite{CarParrinello:1985}. 

However, within DFT, the accurate description of redox reactions remains a challenge. {\color{black} One difficulty steams from the fact that a formal definition of oxidation state is lacking.} The generally accepted IUPAC definition states~\cite{McNaught:1997}: "The oxidation state of an atom is the charge of this atom after the ionic approximation of its heteronuclear bonds". While the charge is often used as a descriptor for the oxidation state in atomistic simulations (e.g., in Bader's~\cite{Bader:1990} and Voronoi's~\cite{Bickelhaupt:1996} methods), it has been shown that this can sometimes lead to misleading descriptions~\cite{Raebiger:2008, Resta:2008}. The challenge of accurately and clearly defining oxidation states has driven the development of novel methods for calculating them from first principles.
For instance, the oxidation state can be defined using topological considerations~\cite{Rappe:2012, Grasselli:2019} based on the modern theory of polarization~\cite{Resta:2007, Vanderbilt:2018}, or it can be determined from the electronic population derived from the eigenvalues of the atomic occupation matrix obtained by projecting the electronic wavefunctions onto specific electronic manifolds (e.g., $d$ orbitals of transition-metal elements)~\cite{Sit:2011}. Whereas the OS as defined in Ref.~\cite{Rappe:2012} has proven to be effective for transport processes~\cite{Pegolo:2020}, the method of Ref.~\cite{Sit:2011} is particularly well suited for the battery cathode materials, as shown in Ref. ~\cite{Timrov:2022b}.

Alongside this general aspect, it is important to note that standard DFT, which uses approximated exchange-correlation functionals, is affected by self-interaction errors (SIEs), leading to unphysical delocalizations of electrons. This limitation prevents accurate modeling of processes where changes in atomic oxidation states are crucial, particularly in systems with strongly localized $d$ or $f$ electrons. Such systems, including {\color{black} transition-metal oxides}, are often of considerable interest for practical applications, such as in rechargeable battery technology~\cite{Hautier:2011, Hautier:2011b}.
Various methods have been proposed to correct or alleviate SIEs: DFT+$U$~\cite{anisimov:1991, Liechtenstein:1995, Anisimov:1997, Dudarev:1998} along with the DFT+$U$+$V$ extension \cite{Campo:2010, TancogneDejean:2020, Lee:2020}, meta-GGA functionals, such as SCAN and other derivatives \cite{Sun:2015,Bartok:2019, Furness:2020} (as well as SCAN+$U$ \cite{Gautam:2018, Long:2020, Kaczkowski:2021, Artrith:2022}), or hybrid functionals (e.g. PBE0 \cite{Adamo:1999} and HSE06 \cite{Heyd:2003, Heyd:2006}), to name a few. {\color{black} In this work, we focus on DFT+U+V, which we will analyze in detail throughout the article, due to its accuracy and affordable computational cost.
In this regard we note how it has recently been shown to improve the prediction of electrochemical and thermodynamic properties when localization is accompanied by significant hybridization, as in Ref.~\cite{Cococcioni:2019}.
Following} the methodology suggested in Ref.~\cite{Sit:2011} to compute oxidation states, {\color{black} DFT+U+V has been shown to deliver sharp ("digital") changes } in oxidation states for transition metals within intercalation cathode materials—such as Mn and Fe phospho-olivines (LMPO, LFPO)—as lithium concentration varies, accurately mirroring the shifts in oxidation states of the transition-metal atoms~\cite{Timrov:2022b}.
 
Thus, applying DFT+U+V to phospho-olivine cathodes provides {\color{black} an ideal starting point to investigate the landscape of oxidation states into material and its evolution.} In this study, we specifically explore finite-temperature effects in LMPO using DFT+U+V FPMD, where, during the dynamics, we observe a clear evolution of {\color{black} OSs} and can track the adiabatic ground state across the redox-active elements. {\color{black}While electron-transfer reactions are most often non-adiabatic processes~\cite{Sit:2006} and can, for example, be described by the general framework of Marcus theory~\cite{Newton:1984, Marcus:1985, Warshel:1991}, it is essential to be able to capture faithfully first the adiabatic ground state driven by the new atomic arrangements resulting from the FPMD evolution.}

FPMD simulations with DFT+U+V offer significant advantages, such as reduced computational cost and enhanced accuracy, compared to hybrid functionals, especially for transition-metal oxides~\cite{Timrov:2022b}. \textcolor{black}{
A key reason is that the screened exchange interaction, captured by the Hubbard parameters, is inherently material-specific in the linear-response formulation of DFT+U+V while hybrid functionals, in addition to cost and complexity, adopt a general recipe that is not always optimal depending on the material or system of interest.}

{\color{black} We showcase here the significant potential of DFT+U+V FPMD; however, the cost of first-principles simulations remains a limiting factor, potentially restricting studies to length and timescales that may be insufficient to observe new phenomena such as ion migration, phase transformations, and chemical reactions.}
{\color{black} Nowadays, it is well established that machine-learning interatomic potentials can help with these challenges. These potentials use a limited set of first-principles calculations for training and are capable to deliver predictions with first-principles accuracy at a cost comparable to that of classical force field simulations~\cite{Behler:2007}.} 
A wide variety of methods exist for building machine learning interatomic potentials; recently, graph neural networks that incorporate equivariant symmetry constraints have been shown to achieve state-of-the-art performance, and notable examples include NequIP~\cite{Batzner:2022} and MACE~\cite{Batatia:2022}. The equivariant nature of these models allows them to outperform some invariant models (e.g., DeepMD~\cite{Zhang:2018}) using a fraction of the data for training. 
However, despite their success, there is still a limiting factor, i.e., lack of control over the atomic {\color{black} OSs}. 
{\color{black} The importance of the atomic OSs stems from the fact that ions with different OS behave differently. This is generally true, but let us consider the example of Mn we will also encounter later:} high spin Mn$^{4+}$ is a non-bonding spherical ion that almost always adopts octahedral coordination with oxygen atoms, Mn$^{3+}$ is a Jahn-Teller active ion that radically distorts its environment, while Mn$^{2+}$ preferably exists in tetrahedral coordination~\cite{Reed:2004}. 

Several schemes have been proposed to describe systems with multiple charge states at the same geometry. Models based on charge equilibration approaches~\cite{Rappe:1991, Ghasemi:2015, Ko:2021, Staacke:2022, Shaidu:2024, Chen:2023}, are expected to incorporate information about oxidation states; however, whether and to what extent they actually capture this information and effectively describe redox chemistry remains an open question, despite recent rationalizations that have offered some insight~\cite{kocer:2024}. Most of these models adopt a dual-learning approach, requiring additional neural networks to learn atomic charges alongside energies and forces. Other models, such as that in Ref~\cite{Deng:2023}, function as universal charge-informed potentials but depend on additional training variables, such as magnetic moments.
Furthermore, some models incorporate the concept of oxidation state through geometrical considerations~\cite{Eckhoff:2020}. While the atomic environment and geometry may at times be sufficient to determine the OS of an atom, this is not universally the case. In liquid environments, for example, the mobility of solvent molecules and other chemical species prevents the establishment of a fixed, well-defined atomic environment that could reliably serve as a label for defining the oxidation state.
Therefore, a simple model oxidation-state informed that could leverage recent state-of-the-art tools (e.g., NequIP~\cite{Batzner:2022} or MACE~\cite{Batatia:2022}), requiring an uncomplicated training procedure without additional dual-learning schemes or training variables, and capable of going beyond information deduced solely from geometry, would represent a valuable option in many cases of study.

{\color{black} In this work, building on the fact that different oxidation states of transition-metal ions behave differently from each other as they were different elements, and emphasizing the accuracy over oxidation states achieved through DFT+U+V, we propose an equivariant neural network potential that treats atoms of the same element with different oxidation states as different species. Once the potential has been generated, we demonstrate not only its accuracy but also that the correct arrangement of oxidation states can be determined through a combinatorial search for the lowest-energy configuration, reproducing correctly the adiabatic re-arrangement of OSs observed in the FPMD. }

\section*{RESULTS}

\subsection*{Static DFT+U+V: oxidation states and voltages}

DFT+U+V is briefly outlined in the Methods section. Here, we discuss the main results where we investigate the phospho-olivine cathode material Li$_x$MnPO$_4$ comparing standard DFT and DFT+U+V. The crystal structure is orthorhombic at $x=0$ and $x=1$ with a $Pnma$ space group. In the present simulations the unit cell contains four formula units, i.e. 24 atoms for $x=0$ and 28 atoms for $x=1$. The four Mn atoms are each coordinated by six oxygen atoms, forming a MnO$_6$ octahedra with the Mn atom centrally positioned. In the DFT+U+V framework the U correction is applied to the $3d$ orbitals of Mn, while the V is set between the $3d$ orbitals of Mn and the $2p$ orbitals of the surrounding O atoms.
The phospho-olivines are known to be antiferromagnetic: we use the magnetic configuration that minimizes the total energy (labeled AF$_1$ in Ref.~\cite{Cococcioni:2019}). We consider all possible concentrations of Li $x$=0, 1/4, 1/2, 3/4, 1.

Using the method described in Ref.~\cite{Sit:2011}, it is possible to determine the OS of Mn atoms within the system by analyzing the electronic population of their $3d$ shells. According to this approach, the occupation numbers of the $3d$ shells are derived from the eigenvalues of the site-diagonal atomic occupation matrix (i.e., $I=J$ in Eq.~\ref{eq:occup_uv}), which has a $5 \times 5$ size in the respective spin-up and spin-down channels. Hence, this approach provides the 10 electronic occupation numbers for the $3d$ shells of Mn. Next, we count how many $d$ states are "fully occupied" (i.e., states with corresponding occupation number approximatively 1) and compare to the valence electronic configuration of the Mn atom, which has 7 valence electrons: $3d^5 4s^2$.
For example, in the fully de-lithiated case with $x=0$, the DFT+U+V calculation provides the following occupations for all four Mn atoms: 
$n_i^{\uparrow} = 0.50$, $\textbf{0.99}$, $\textbf{0.99}$, $\textbf{1.00}$, $\textbf{1.00}$ for the spin-up channel, and $n_i^{\downarrow} = 0.05$, $0.06$, $0.08$, $0.09$, $0.22$ for the spin-down channel. Thus, the fully occupied states (indicated in bold) are 4; this indicates that 4 electrons are assigned to each Mn atom in the compound. Since Mn has 7 valence electrons, this implies that 7–4=3 electrons are involved in bonding with the oxygen environment, yielding an oxidation state of 3+. The fractional occupations of nominally empty states are a result of hybridizations with O-$2p$ states that contribute to the projections.  
Furthermore, greater deviations from 0 in the eigenvalues signal stronger mixing of unoccupied $d$ orbitals with ligand orbitals, as illustrated by the first spin-up eigenvalue. Now, let us consider the fully lithiated case with $x=1$, where the system is fully lithiated (with 4 additional Li$^+$ ions and 4 extra electrons). For each Mn atom, our DFT+U+V calculation provides the following occupations: $n_i^{\uparrow} = \textbf{0.99}$, $\textbf{0.99}$, $\textbf{1.00}$, $\textbf{1.00}$, $\textbf{1.00}$ and $n_i^{\downarrow} = 0.02$, $0.02$, $0.03$, $0.07$, $0.08$. The fully occupied states (in bold) indicate that each Mn atom now accommodates 5 electrons, resulting in an oxidation state of 2+: each Mn atom has gained an electron and has been reduced. 

\begin{figure}
    \centering
    \includegraphics[width=0.8\textwidth]{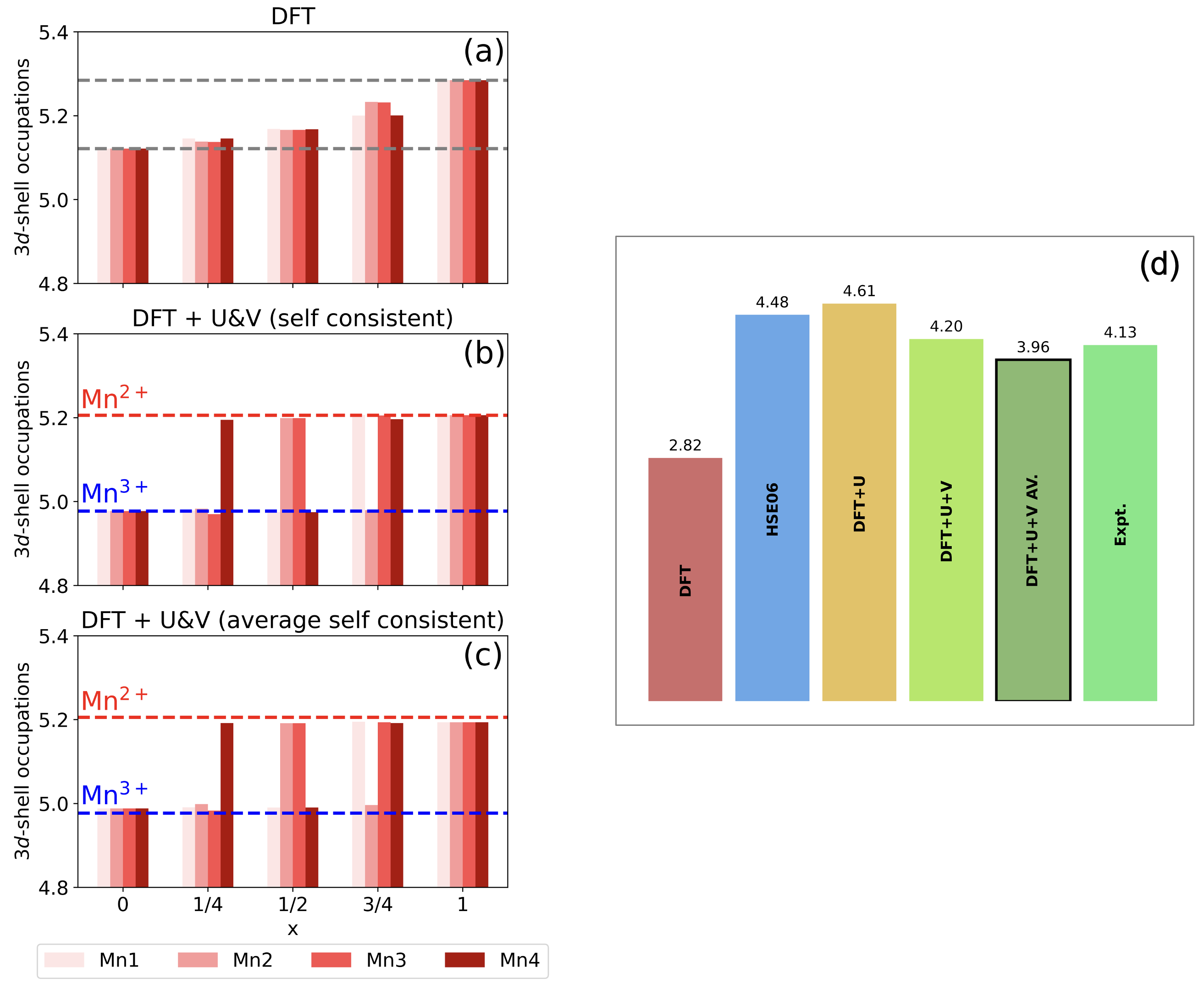}
    \caption{\textbf{(a-c)} Löwdin occupations $n$ (sum of electronic occupation numbers) of the Mn 3d shells in Li$_x$MnPO$_4$ at x=0, 1/4, 1/2, 3/4, 1, calculated using three approaches: (a) DFT, (b) DFT+U+V with self-consistent U and V, and (c) DFT+U+V with average parameters (U = 5.1 eV, V = 0.7 eV). Each bar represents the Löwdin occupation of one of the four Mn atoms in the unit cell, distinguished by a different shade. In panels (b) and (c), the red horizontal line indicates $n=5.21$ (Mn$^{2+}$), while the blue line indicates $n=4.98$ (Mn$^{3+}$). \textbf{(d)} Voltages (in V) for Li$_x$MnPO$_4$, computed using DFT+U+V with averaged self-consistent U and V, compared with DFT, HSE06, DFT+U, and DFT+U+V with self-consistent U and V~\cite{Timrov:2022b}. Experimental data are from Refs.~\cite{Muraliganth:2010, KOBAYASHI2009}.}
    \label{fig:os_volt}
\end{figure}

We now move on to examining the system's behavior at intermediate Li concentrations. For simplicity, we can use as a descriptor of oxidation state the sum of the electronic occupations $n = \sum_{i} (n_i^{\uparrow} + n_i^{\downarrow})$. This quantity, often referred to as \text{Löwdin occupation}, is particularly useful for bookkeeping~\cite{Raebiger:2008} and especially for describing the (de-)lithiation process, as will be discussed below. 
{\color{black} For example, applying this definition to the DFT+U+V $3d$-shells occupations of Mn presented above we find $n = 4.98$ and $n = 5.21$ for the fully de-lithiated and fully lithiated systems, respectively.}
In Fig.~\ref{fig:os_volt} (a-c), we present the L\"owdin occupations of the $3d$ shells of Mn atoms in Li$_x$MnPO$_4$ for intermediate Li concentrations (i.e., at $x = 0$, $1/4$, $1/2$, $3/4$, $1$), computed using three different approaches: (a) standard DFT, (b) DFT+U+V with self-consistent U and V, and (c) DFT+U+V with U and V parameters averaged over all Mn sites and all possible Li concentrations. The four Mn atoms within the unit cell are represented by bars in different shades. In going from $x=0$ to $x=1/4$, one Li atom is added to the system, which introduces one Li$^+$ cation and one electron. Then, additional Li atoms are incrementally introduced until $x=1$ is reached with four Li$^+$ cations and four additional electrons in the system. Fig.~\ref{fig:os_volt} (a) shows that standard DFT delocalizes every electron added across all Mn sites, resulting in no significant differences in the L\"owdin occupations across sites as the Li concentration changes; no individual Mn ion undergoes a redox reaction. Fig.\ref{fig:os_volt} (b) presents the same study using DFT+U+V, and calculating self-consistent U and V parameters. 
To distinguish more clearly the oxidation states of Mn, we indicate with a blue dashed horizontal line the occupation level corresponding to the concentration $x=0$, where all Mn atoms are 3+ ($n=4.98$), and with a red dashed line the corresponding occupation level for the concentration $x=1$, where all Mn atoms are 2+ ($n=5.21$). 
DFT+U+V shows a clear "digital" change in the Löwdin occupations: during the lithiation process, the addition of one Li$^{+}$ ion and one electron to the cathode alters the occupation of a single Mn ion from 4.98 to 5.21, corresponding to a change in oxidation state from 3+ to 2+, while all other Mn ions remain unaffected. This process continues with further Li intercalations, ultimately resulting in the reduction of all Mn ions from 3+ to 2+. Consequently, DFT+U+V with self-consistent parameters accurately captures the mixed-valence character of the Li$_x$MnPO$_4$ compound, which includes two distinct Mn ion states, Mn$^{3+}$ and Mn$^{2+}$, at $x= 1/4$, $1/2$, $3/4$. In contrast, standard DFT fails to localize the additional electrons on a specific Mn ion; instead, the charge density is delocalized, spreading almost equally across all Mn ions, resulting in approximately equal occupations, as shown in Fig.~\ref{fig:os_volt} (a). {\color{black} Thus, in DFT, at $x = 1/4$, $1/2$, $3/4$, there is effectively a single type of Mn ion with intermediate occupation values that progressively shift with Li content, corresponding to "unchemical" oxidation states of Mn$^{2.25+}$, Mn$^{2.5+}$, Mn$^{2.75+}$.}

We now want to explore the use of the DFT+U+V at finite temperatures by means of FPMD. During MD simulations, the positions of the atoms change, which can lead to variations in the Hubbard parameters. Ideally, U and V should be re-evaluated from first principles for each MD frame (self-consistently at fixed geometry). However, it is currently impractical to add the computationally intensive calculation of Hubbard parameters to each frame of an already demanding FPMD simulation. {\color{black} To address this challenge, machine learning approaches employed to predict the Hubbard parameters will greatly help~\cite{Uhrin:2024}.}
Here, we examine whether the use of average U and V parameters, kept fixed during the MD runs, can still describe accurately the "digital" change of OSs of Mn atoms shown in Fig.~\ref{fig:os_volt} b. These U and V parameters are calculated as the average of the self-consistent values (see Supplementary Table 1) across all Li concentrations, resulting in U = 5.1 eV and V = 0.7 eV. The results are shown in Fig.~\ref{fig:os_volt} c. 
In order to quantify the effect of using average self-consistent U and V as opposed to the fully self-consistent ones, we display the same L\"owdin occupation treshold for Mn$^{2+}$ ($n=4.98$, red dashed line) and Mn$^{3+}$ ($n=5.21$, blue dashed line) as obtained with the self-consistent parameters shown in Fig.~\ref{fig:os_volt} b. Clearly, even when using average parameters the digital variations in the OSs of Mn atoms are preserved. The only effect is a slight reduction in the contrast between the 2+ and 3+ levels. 

 {Finally, in Fig.\ref{fig:os_volt} (d) we report the voltages $\Phi$ in comparison with different methods~\cite{Timrov:2022b} and experiments~\cite{Muraliganth:2010, KOBAYASHI2009}. The voltages are computed following the relation~\cite{Cococcioni:2019} $-e\Phi=E_{x=1}-E_{x=0}-E_{Li}$ where $-e$ is the electron charge, $E_{x=1}$ and $E_{x=0}$ are the total energies per formula unit of Li$_x$MnPO$_4$ at $x=1$ and $x=0$, respectively. $E_{Li}$ is the total energy of bulk Li (representing the anode). }
The  voltage computed with average self-consistent U and V is 3.96 V, slightly underestimating the one computed with self-consistent U and V, 4.20 V. Nevertheless it is still closer to the experiment than standard DFT, hybrid functionals (HSE06), and DFT+U (for the voltage estimate we have re-optimized the cell and the atomic positions with the new set of average U and V). 

In summary, the study of average U and V parameters across oxidation states and voltages supports their applicability as fixed parameters throughout the DFT+U+V FPMD simulations, with details and further investigations presented in the following section.

\subsection*{DFT+U+V molecular dynamics}

We performed FPMD on the Li$_x$MnPO$_4$ system, where energies and forces are calculated using DFT+U+V (Eq.~\ref{E_uv} and Eq.~\ref{forces}) with average U and V parameters (U=5.1 eV and V=0.7 eV). Simulation were carried out for all possible Li concentrations: $x= 0$, $1/4$, $1/3$, $3/4$, $1$. The simulations were performed with a time-step $\delta t = 4$ fs within the NVT ensemble by using the stochastic-velocity rescaling algorithm~\cite{Bussi:2007}. For each Li concentration, we performed two simulations at different temperatures: T=400 K and T=900 K. In passing, we note that a higher temperature is necessary to allow Mn atoms to explore all possible configurations of their {\color{black} OSs} (i.e., all possible combinations) {\color{black} in the timescale of the simulations.}

\begin{figure}
    \centering
    \includegraphics[width=0.72\textwidth]{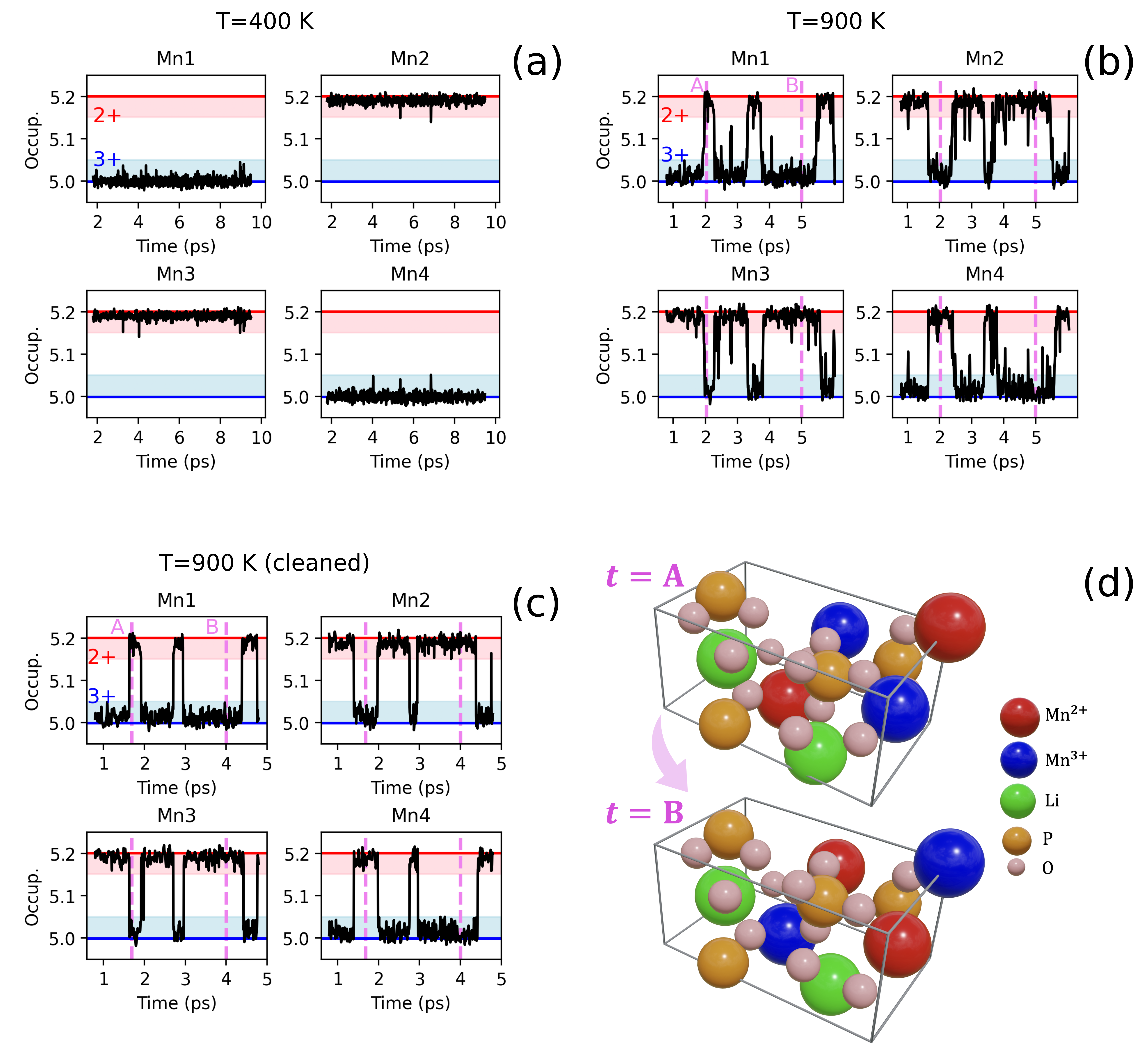}
    \caption{\textbf{DFT+U+V ab-initio molecular dynamics of Li$_x$MnPO4 (x=1/2) in the NVT ensemble.} \textbf{(a)-(c)} Time evolution of the 3d-shell Löwdin occupations $n$ (Occup.) for the four Mn atoms. Red and blue horizontal lines indicate static Löwdin occupations for Mn2+ ($n=4.98$) and Mn3+ ($n=5.21$), respectively, while the light red and light blue areas highlight the range of fluctuations within the respective oxidation states. DFT+U+V FPMD is performed at \textbf{(a)} T=400 K, and \textbf{(b)} T=900 K. \textbf{(c)} shows the same as in (b) but excluding configurations with intermediate occupations (outside the fluctuation range). \textbf{(d)} Two frames of the dynamics showing different oxidation state patterns of Mn. Mn$^{2+}$ and Mn$^{3+}$ are shown in red and blue, respectively, Li in green, O in pink, and P in orange. They represent the configurations indicated in (b) and (c) with vertical violet dashed lines: configuration A (top) and configuration B (bottom).}
    \label{fig:md}
\end{figure}

In Fig.~\ref{fig:md}, we show an example of the evolution of the Löwdin occupations over time for the four Mn atoms, for the system with Li concentration $x=1/2$, consisting of two Li$^+$ ions and two additional electrons that can localize on any Mn atom. To facilitate the analysis, we include the occupation levels previously discussed in the static case to distinguish the two {\color{black} OSs} of Mn: the 3+ levels ($n=4.98$) indicated by the blue line and the 2+ levels ($n=5.21$) indicated by the red line. In the simulation at T=400 K (Fig.~\ref{fig:md} a), the occupations fluctuate within a small range of values but remain close to the initial level associated with the specific OS. Specifically, Mn1 and Mn4 remain in the 3+ state, while Mn2 and Mn3 remain in the 2+ state. The situation changes when we consider the temperature T=900 K (Fig.~\ref{fig:md} b). The evolution of the occupations reveals distinct jumps between the two OSs. Electrons are exchanged adiabatically between Mn atoms, so there are always two Mn$^{2+}$ and two Mn$^{3+}$. For example, we see that approximately in the first 2 picoseconds, Mn1 is 3+, and correspondingly, Mn3 is 2+. Subsequently, we observe a sharp shift in OSs: Mn1 switches to the 2+ state, while simultaneously, Mn3 shifts to the 3+ state, indicating that the electron has moved adiabatically from Mn3 to Mn1. As can be observed, such transitions in OSs continue over time and also involve the other two Mn atoms. In the Supplementary Material, we further discuss this point, also examining the individual electronic occupations; {\color{black} importantly, we reitarate that this is not intended to reproduce the physical dynamics of the polaron hopping and tunneling but just the exploration of the electronic ground states.}

The finite temperature causes small fluctuations in the occupations, during which the atoms reasonably keep their OSs. However, when the new atomic configuration in the dynamics favors a different OS pattern, sharp transitions occur, which are distinct and well separate from the thermal fluctuations. 
In these figures, the extent of the fluctuations is represented by light red and light blue bands, within which Mn$^{2+}$ and Mn$^{3+}$ atoms preserve their own OS.

In Figs~\ref{fig:md} (a) and (c), two specific configurations A and B are highlighted (with vertical violet dashed lines) where there is a change in the distribution of OSs. In A, the four Mn atoms—Mn1, Mn2, Mn3, and Mn4—are 2+, 3+, 3+, and 2+, respectively. In B they are are 3+, 2+, 2+, and 3+, respectively. As an example, in Fig.~\ref{fig:md} (d), the corresponding atomic configurations are shown. Mn$^{2+}$ atoms are red, Mn$^{3+}$ blue, and they are surrounded by other LMPO atoms, depicted in different colors. The upper cell shows configuration A, while the lower one shows configuration B, visually highlighting how, in these particular example, the Mn atoms make a transition with exactly opposite OSs.

In the following section, we will explain how we selected data from the FPMD simulations to train the machine learning potential. Then, we test the potential and investigate its predictive power in determining the atomic OSs of Mn atoms on specific configurations.

\subsection*{Neural network potential and oxidation state identification} 

The FPMD trajectories are utilized to train an equivariant NN potential. For this purpose, we employ NequIP, which has demonstrated state-of-the-art performance~\cite{Batzner:2022}. 
The training and validation datasets are obtained following the procedure described below (a schematic representation of the workflow is also provided in Supplementary Figure 1).
DFT+U+V FPMD simulations are conducted for each x$_{Li}$ concentration of Li at temperatures T=400 K and T=900 K. Each trajectory spans approximately 9 ps, yielding a total combined trajectory length of around 92 ps. The trajectories are filtered as shown in Fig.~\ref{fig:md} (c), ensuring that all Mn atoms have occupations within the red and blue bands ---that is, excluding configurations where at least one Mn atom has L\"owdin occupations $n$ $ 5.05 \leq n \leq 5.15 $ out of the average fluctuations associated to each OS. {\color{black} After this filtering, we obtain training segments summing up to 80 ps.} 
Then, we proceed to identify all possible OSs patterns; those depend on the Li-ion concentration, which determines the number of additional electrons and in turn the number of Mn$^{2+}$ atoms. Furthermore, for a fixed Li-ion concentration (i.e., at fixed number of Mn$^{2+}$ and Mn$^{3+}$), the OSs patterns of the Mn atoms can vary, as the different atomic configurations in the molecular dynamics simulations may promote the redistribution of localized electrons across Mn sites.
Let us consider the example of x$_{Li}=1/4$ with one Li$^{+}$ cation and one additional electron in the system (with respect to the fully delithiated structure where all Mn atoms are 3+). The electron localizes on one of the Mn atoms, shifting its OS from Mn$^{3+}$ to Mn$^{2+}$, while the others remain Mn$^{3+}$. Since any of the four Mn$^{3+}$ atoms can receive the electron and become Mn$^{2+}$, there are four possible OSs patterns. For simplicity, if we represent Mn$^{2+}$ as 0 and Mn$^{3+}$ as 1, the four patterns correspond to the permutations of 4 elements with 3 repeated (i.e., $4!/3!=4$): 0111, 1011, 1101, and 1110, where each number represents, in order, the OS of Mn1, Mn2, Mn3, and Mn4. Similarly, the concentration x$_{Li}=3/4$ includes 4 possible OSs patterns, having now three Mn$^{2+}$ and one Mn$^{3+}$: 1000, 0100, 0010, 0001. For the concentration $x_{Li}=1/2$, there are (4!/{2!/2!}=6) 6 possible OSs patterns: 0110, 0011, 0101, 1001, 1010, 1100. Lastly, at x$_{Li}=0$, all Mn atoms are in the 3+ state (1111), while at x$_{Li}=1$, all Mn atoms are in the 2+ state (0000): each contributing with a single OS pattern. Thus, by adding up all possible OSs patterns for all Li-ion concentrations, we get a total number of 16 possible OSs patterns. In our simulations, we find that the FPMD at T=400 K spans 9 possible of them across all concentrations, while the FPMD at T=900 K includes the whole set of 16 patterns, although some correspond to a very small number of configurations. For example, the patterns 1010, 0011, and 0101 appear in only 31, 44, and 13 snapshots, respectively. 

Then, each atomic configuration is assigned to a particular pattern of OSs and this information is used to build the ensemble of data for the training and validation of the neural network potential. We proceed as follows (see also Supplementary Figure 1): for each temperature and OS pattern we select up to a maximum of 100 snapshots (if available), randomly chosen in order to minimize correlations as much as possible in the available trajectories. As a result of this procedure we automatically include all possible Li concentrations. This selection ends up with a total number of 2288 snapshots that are divided into a training (80 \%) and a validation set (20 \%), to monitor training over time.        

\begin{figure}[t]
    \centering
    \includegraphics[width=0.85\textwidth]{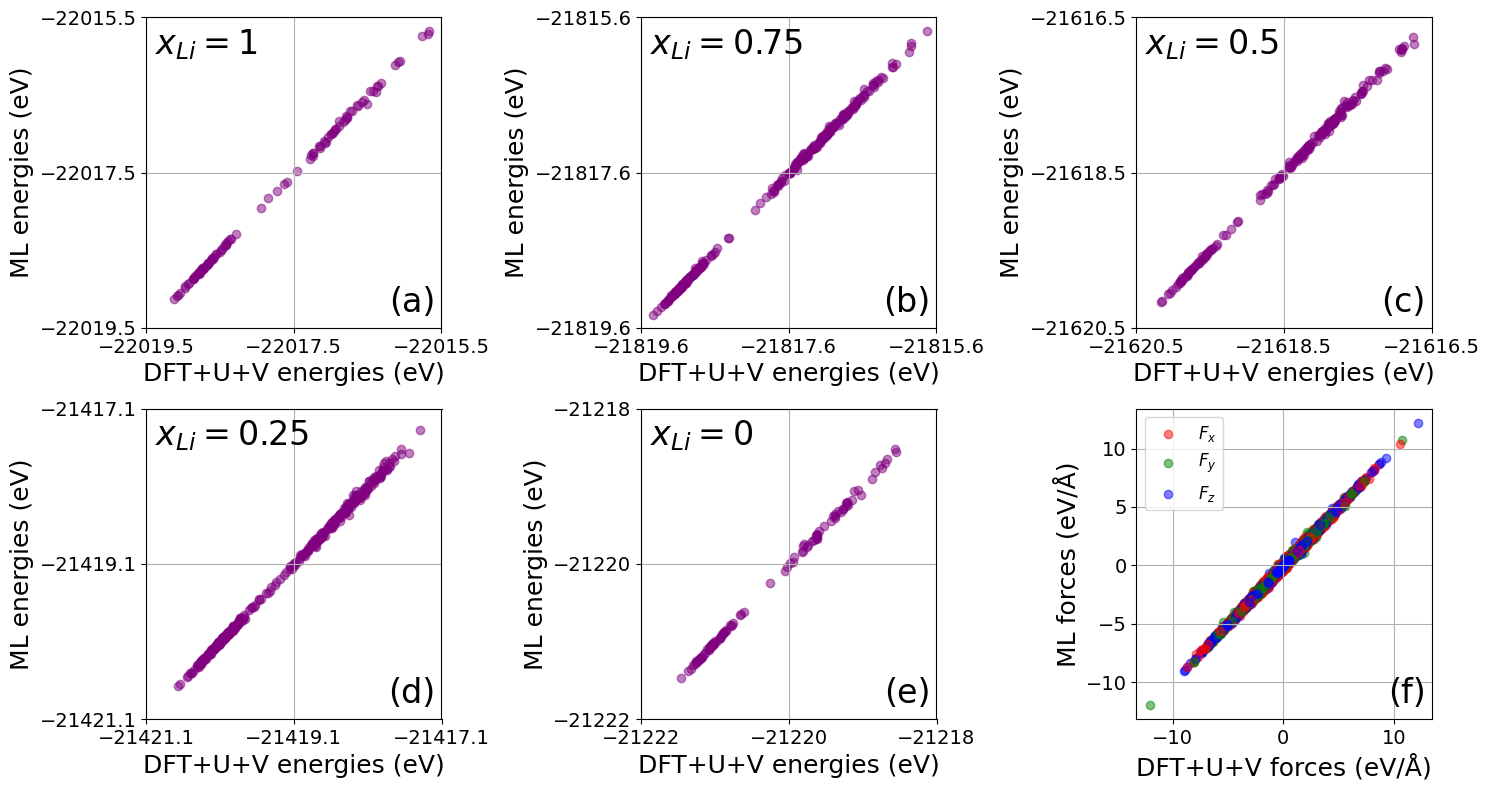}
    \caption{\textbf{Parity plots}. \textbf{(a)-(e)} Machine learning predictions of total energies vs. DFT+U+V total energies for all x$_{Li}$ concentrations of Li considered in the FPMD. \textbf{(f)} Machine learning predictions of forces vs. DFT+U+V forces (the Cartesian components are displayed with different colors). Each plot shows linear correlation coefficient $\sim$0.995.}
    \label{fig:pplot}
\end{figure}

As mentioned, we train an equivariant neural network potential by using NequIP~\cite{Batzner:2022} and using atomic positions and DFT+U+V forces; Mn in different OSs are defined as different atomic types. 
{\color{black} Then, for testing, we select a maximum number of 50 random frames (if available) for each OS pattern and temperature, not included in the training and validation set.} This results in an ensemble of approximately 1000 frames with which we perform error analysis and evaluate performance of the model. The mean absolute error (MAE) on energies is 16.6 meV, the MAE on energies per atom is 0.64 meV, and the MAE on forces is 41 meV/Å. {\color{black} Then, the same testing data-set is used for parity plots} in Fig.~\ref{fig:pplot}, where we compare the machine learning predictions of energies and forces vs. the DFT+U+V corresponding results reported in Fig.~\ref{fig:pplot}. The total energies are naturally clustered, each cluster associated to a different Li concentration. Both studies on energies and forces show excellent linear correlation (linear correlation coefficient $r \sim 0.995$ in all cases).

\begin{figure}
    \centering
    \includegraphics[width=0.8\textwidth]{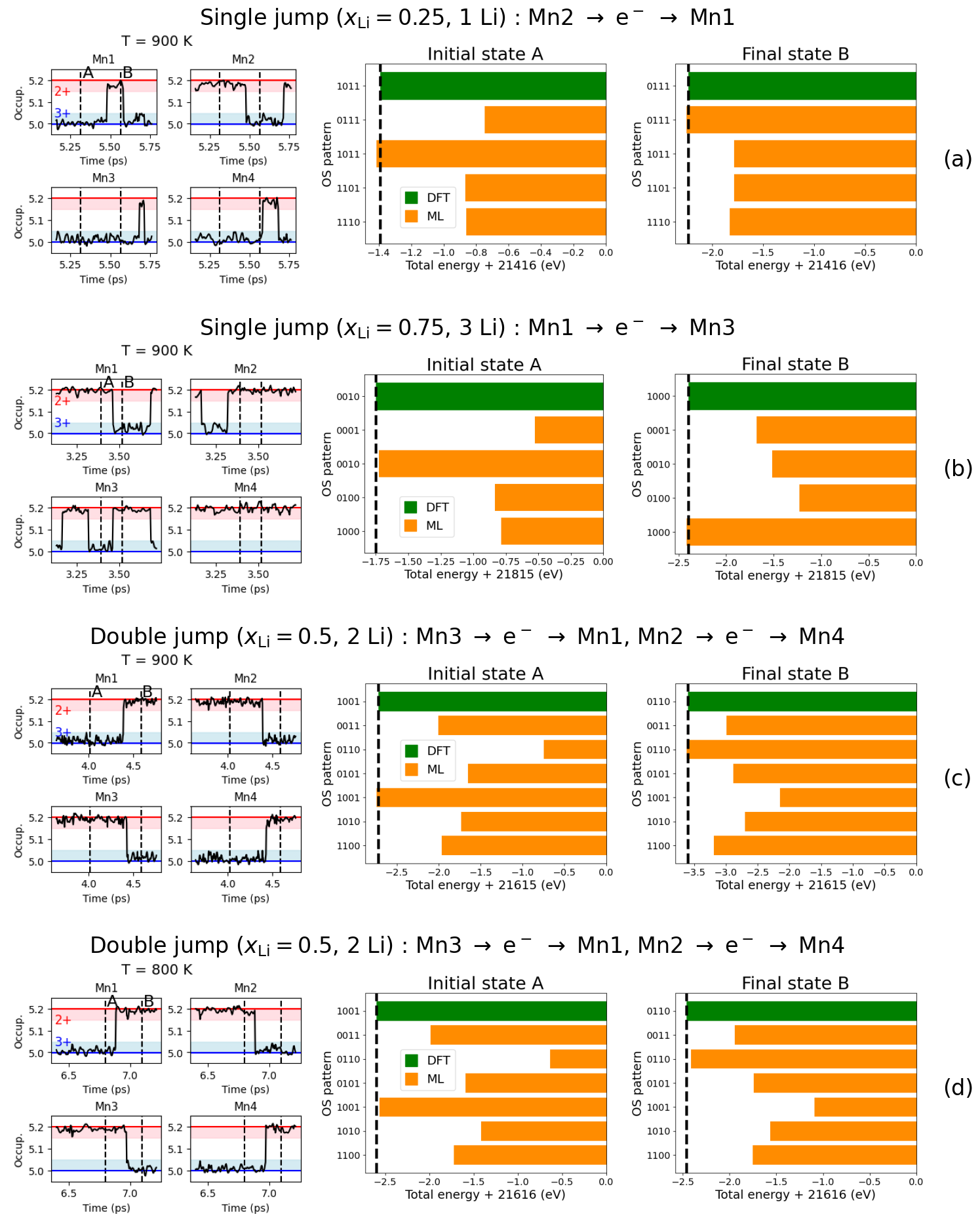}
    \caption{\textbf{Oxidation states re-arrangements of Mn atoms in the DFT+U+V FPMD of Li$_x$MnPO$_4$}. 
    \textbf{Left panel}. Time evolution of $3d$-shells L\"owdin occupations (Occup.) for the four Mn atoms in the unit-cell. Horizontal red and blue lines (and corresponding shaded areas) represent the limit associated to Mn$^{2+}$ and Mn$^{3+}$, respectively. A and B (vertical black dashed lines) indicate two configurations (out of the training set) with a different OS pattern.
    \textbf{Central and right panels}. DFT+U+V energy (green) compared to the ML energy predictions (orange) of all possible oxidation states patterns for fixed atomic configurations A and B. 
    }
    \label{fig:test_nn}
\end{figure}

As described above, during the FPMD simulations, Mn atoms change their OSs in response to the re-arrangement of the atomic positions in the system.
We now conduct an in-depth analysis of the ability of the neural network potential to predict such OSs transitions. The study is presented in Fig.~\ref{fig:test_nn} and described in the following. 
First, only configurations outside the training and validation dataset are chosen. In Fig.~\ref{fig:test_nn} (a), we consider a segment of the FPMD trajectory of Li$_x$MnPO$_4$ with $x=1/4$ at T=900 K. At this concentration, we have a single Li$^{+}$ ion and one additional electron that localizes on one Mn atom. Then, we consider two atomic configurations, A and B, which are separated by an adiabatic "jump" of the electron from one Mn to another, indicated in the figure by dashed black vertical lines. For configuration A, the electron is on the Mn2 atom, which is 2+, while all the other Mn atoms are 3+. For configuration B, the electron has switched to the Mn1 atom, which now is the one with OS 2+. We refer to this re-arrangement of OSs as "single jump", meaning that only one couple of Mn atoms and one electron are involved, and we denote it as Mn$2$ $\xrightarrow[]{} e^{-} \xrightarrow[]{}$ Mn$1$. The OSs patterns of Mn atoms in configuration A is 1011, and in configuration B is 0111.
On the right side of Fig.~\ref{fig:test_nn} (a) we report block diagrams comparing DFT+U+V total energies of these configurations A and B (green) with ML predictions (orange) for all possible 4 OSs patterns. 
The study reveals that the DFT+U+V FPMD configuration is accurately identified as the one with the lowest energy by the machine-learning combinatorial search. This occurs for both the initial and final states, A and B. Thus, remarkably, the methodology precisely predicts the correct pattern of OSs.

In Fig.~\ref{fig:test_nn} (b), we select instead a FPMD trajectory at 900 K with a Li concentration $3/4$, meaning there are three additional electrons in the system and, as before, still 4 OSs patterns. The selected configurations A and B still lead to a "single jump" process where the electron moves according to Mn$2$ $\xrightarrow[]{} e^{-} \xrightarrow[]{}$ Mn$1$. Evaluating the various OSs patterns with the neural network potential once again the DFT+U+V one is identified as the one having the lowest energy among all possible combinations. 

In Fig.~\ref{fig:test_nn} (c), a trajectory at a concentration $1/2$ is considered, where there are two additional electrons and 6 possible OSs patterns. The process is now a "double jump", for which, between configurations A and B, electrons move as follows: Mn$3$ $\xrightarrow[]{} e^{-} \xrightarrow[]{}$ Mn$1$ and Mn$2$ $\xrightarrow[]{} e^{-} \xrightarrow[]{}$ Mn$4$. Once again, the machine-learning potential identifies the correct OS pattern for both the initial state A and the final state B, accurately reproducing the rearrangement of OSs. 

Last, we performed a new, independent DFT+U+V FPMD simulation at a Li concentration of x=1/2 and a different temperature of T=800 K. In Fig.\ref{fig:test_nn} (d), we show a portion of this new trajectory, which is completely independent of the previous datasets. We observe a "dual jump" process similar to that reported in Fig.\ref{fig:test_nn} (c). Remarkably, the exploration and evaluation of various OSs patterns by the machine-learning potential once again proves highly effective in accurately describing the rearrangement of OSs.
\begin{figure}
    \centering
    \includegraphics[width=1.\textwidth]{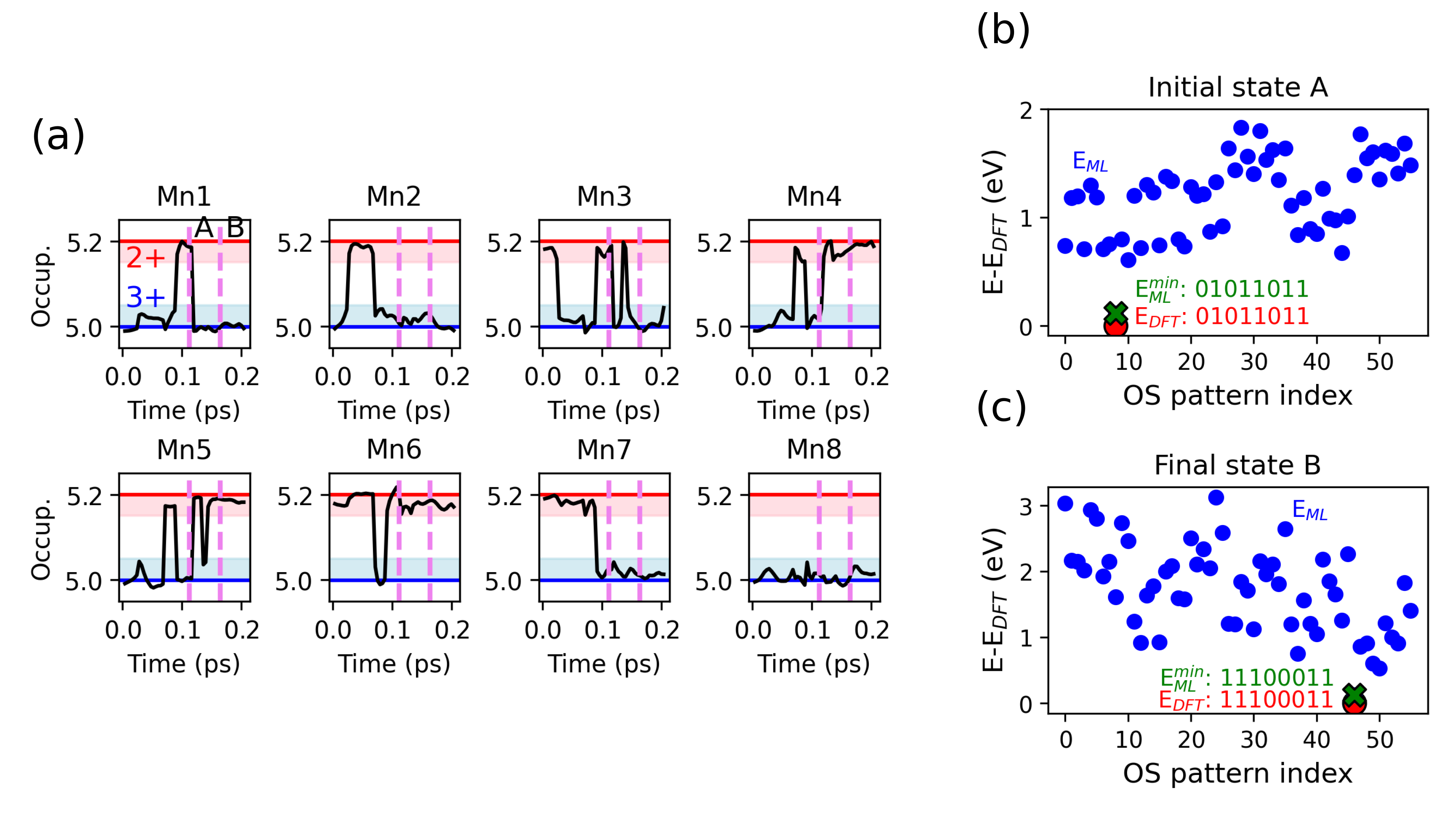}
    \caption{\textbf{Oxidation states rearrangements of Mn atoms in the DFT+U+V FPMD of Li$_x$MnPO$_4$ in a supercell with x=3/8.} \textbf{(a)} L\"owdin occupations of $3d$-shells of Mn atoms in the supercell. The configurations A e B are indicated in vertical violet dashed lines. From A to B Mn1, Mn3, Mn4 and Mn5 shift their OSs. \textbf{(b)} ML energy predictions ($E_{ML}$) over all possible OS patterns in configuration A (blue points), ML minimum energy $E_{ML}^{min}$(green cross). DFT(+U+V) energy ($E_{DFT}$) of configuration A (red point). \textbf{(c)} The same of (b) but for configuration B. The minimum-energy pattern predicted with ML is indicated and compared with the DFT+U+V result: 0 and 1 mean Mn$^{2+}$ and Mn$^{3+}$, respectively. }
    \label{fig:test_sc}
\end{figure}

In Fig.~\ref{fig:test_sc} we present the same investigation for an even larger supercell. The supercell is built by doubling the fully-lithiated unit cell along the $z$ direction and randomly excluding 5 Li atoms (among the available 8) to obtain the unseen concentration $x=3/8$. The supercell includes thus 51 atoms. A portion of the FPMD trajectory at 800 K is presented in Fig.~\ref{fig:test_sc} (a) where we report the time evolution of the L\"owdin occupations for the eight Mn atoms present in the supercell. Between the configurations A and B (indicated with vertical dashed lines in the figure) four Mn atoms shift their OS, namely Mn1, Mn3, Mn4 and Mn5. In this case, a larger number of OSs patterns are observed, specifically $8! / 5! / 3! = 56$. The energy of all of them is evaluated with the neural network potential for both A and B configurations and the results are shown in Fig.~\ref{fig:test_sc} (b) and (c). 
The pattern of OSs given by DFT+U+V is succesfully predicted by the machine learning as the minimum energy configuration for both A and B states, reproducing the OSs transitions observed in the dynamics. {\color{black} Of course, in much larger supercells, it would not be intended to perform exponentially exploding combinatorial searches, substituting instead local explorations or Monte Carlo moves.}

\section*{DISCUSSION}

In the following, we schematically outline the main findings of this work:

\begin{itemize}

\item The DFT+U+V molecular dynamics approach has proven effective in exploring the ground-state landscape of oxidation states for redox-active elements; here, with the specific example of Mn in the LMPO cathode system. To achieve this, we used fixed U and V parameters derived from the ab-initio self-consistent ones averaging over all Mn atoms and Li concentrations. This was justified by the fact that average U and V still provide electrochemical properties -- e.g., voltages -- in better agreement with experiments than hybrid functionals, as well as a sharp description of oxidation states transitions.
In passing, we add that DFT+U+V molecular dynamics is much less computationally intensive than hybrid functionals molecular dynamics. So, it is both more convenient and more accurate. If on-the-fly re-optimization of U and V parameters where necessary during the dynamics, machine learning models predicting Hubbard parameters~\cite{Uhrin:2024} may be of great help.
For all these reasons, we believe DFT+U+V FPMD represents the state-of-the-art in descriptions of systems where localization and hybridization of $d$ and $f$ electrons play a central role.

\item We showed how a machine learning potential can be trained to identify the correct oxidation states, and switch those whenever necessary. The training is done on atomic positions and forces, but treating redox-active elements with different oxidation states as distinct types, where the oxidation states are determined by means of DFT+U+V. Notably, this procedure is expected to be agnostic to the method used for determining oxidation states (DFT+U+V) as well as to the specific machine learning machinery employed (NequIP).
Regarding the potential’s training, the inclusion of various Li concentrations plays a crucial role in capturing the range of oxidation state patterns for subsequent predictions. However, it is reasonable to assume that not all concentrations are necessary. Recent studies suggest that including only two non-trivial concentrations may suffice to yield low errors for both energies and forces~\cite{Goodwin:2024}. Nevertheless, further investigations are needed to determine how reducing the number of concentrations might still yield accurate predictions of the oxidation state patterns.
Future developments include automating the process, and ideally coupling it with Monte Carlo algorithms to simplify the combinatorial search of OSs patterns (especially required for supercells where the number of OSs patterns greatly increases). This process could be further streamlined by an initial guess based on the atomic environment geometry of redox-active elements.

We believe this work could greatly enhance predictive, atomistic electrochemistry across multiple fields of science, due to its simplicity and generality. For example, we expect it to be easily extendable to scenarios where geometric information alone is clearly inadequate for determining oxidation states, such as in liquids.

\end{itemize}


\section*{METHODS}
\subsection*{DFT+U+V} \label{sec:dftuv}
The DFT+U+V method was originally introduced in Ref.~\cite{Campo:2010} as a generalization of DFT+U (in its simplest, rotationally invariant formulation introduced by Dudarev et al.~\cite{Dudarev:1998}) and it is based on an extended Hubbard model that contains both on-site U and intersite V electronic interactions. 
The physical rationale for the Hubbard U and V corrections lies in their ability to address spurious deviations from the piecewise linearity (PWL) of the DFT total energy with respect to the fractional addition or removal of charge~\cite{perdew:1982, cococcioni:2002, Cococcioni:2005, Kulik:2006, MoriSanchez:2006, morisanchez:2009, zhao:2016}, which are associated with the self-interaction errors (SIEs).
In DFT+U+V, such deviations from PWL are handled by adding an extended Hubbard corrective term $E_{U+V}$ to the standard DFT energy $E_{DFT}$:
\begin{equation}\label{E_uv}
E_{DFT+U+V} = E_{DFT} + E_{U+V}.
\end{equation}
The extended Hubbard correction energy for a manifold with angular momentum $l$ is written as:
\begin{equation}\label{E_uv2}
E_{U+V} = \sum_{I, \sigma} \frac{U^I}{2} 
Tr[\mathbf{n}_l^{II \sigma} (\mathbf{1} - \mathbf{n}_l^{II \sigma})]-
 \sum_{I,J,\sigma}^{*} \frac{V^{IJ}}{2} Tr [\mathbf{n}_l^{IJ \sigma} \mathbf{n}_l^{JI \sigma}],
\end{equation}
where $I$ and $J$ are atomic site indices, $\sigma$ labels the spin of electrons. $U^J$ and $V^{IJ}$ are effective on-site and inter-site Hubbard parameters, respectively. The asterisk in the sum signifies that, for each atom \textit{I}, the index \textit{J} include all its neighbors up to a given distance. The generalized occupation matrices $\bold{n}_l^{IJ \sigma}$ are derived from the projection of Kohn-Sham (KS) states onto localized atom-centered orbitals $\phi_{m}^{J}$ (Hubbard projector functions):
\begin{equation}\label{eq:occup_uv}
\bold{n}_l^{IJ \sigma} \equiv n^{IJ \sigma}_{m, m'} = \sum_{\mathbf{k},v} f_{\mathbf{k} v}^{\sigma} \braket{ \psi_{\mathbf{k} v}^{\sigma} | \phi_{m'}^{J}}
\braket{ \phi_{m}^{I} | \psi_{ \mathbf{k} v}^{\sigma}},
\end{equation}
where $v$ and $\sigma$ represents the band and spin labels of the KS wavefuncitions $\psi_{\mathbf{k} v}^{\sigma}$, respectively, $\mathbf{k}$ denotes points in the first Brillouin zone, and $f_{\mathbf{k} v}^{\sigma}$ are the occupations of the KS states. We also point out that the choice of the projector function exerts a significant influence on the numerical values of the calculated Hubbard parameters and, more generally, on the prediction of material properties~\cite{Timrov:2020b}.
The on-site $U^{I}$ and intersite $V^{IJ}$ terms work in opposition to each other. The on-site term encourages localization at atomic sites, reducing hybridization with neighboring atoms, while the inter-site term favors the formation of hybridized states involving neighboring atoms. As a result, the values of $U^{I}$ and $V^{IJ}$ play a crucial role in balancing localization and hybridization in Hubbard-corrected DFT. However, since they are not known a priori, they must be determined through an appropriate method. 

While in the simplest cases these parameters could be obtained through semi-empirical tuning (but then negating the predictive power of the approach, and the capability to deal with complex and very diverse local environments, that require atom-specific U and V), unbiased predictions identify Hubbard parameters self-consistently through linear-response calculations~\cite{Cococcioni:2005, Shishkin:2016, Cococcioni:2019}, particularly efficient when density-functional perturbation theory (DFPT) is deployed~\cite{Timrov:2021, Timrov:2018}. 

Moreover, it has been shown that jointly optimizing Hubbard parameters and the crystal structure, rather than relying on the equilibrium geometry obtained using (semi-)local functionals, can significantly improve the accuracy of the final properties of interest~\cite{Timrov:2020b}. To do so, a self-consistent procedure combining DFPT and structural optimizations can be used~\cite{Timrov:2021, Hsu:2009}. 

Finally, we mention that, from the total energy expressed in Eqs.~\ref{E_uv} and~\ref{E_uv2}, it is possible to calculate the extended Hubbard forces. Under the Born-Oppenheimer approximation, the force acting on atom $I$ situated at position $\textbf{R}_I$ is defined as:
\begin{equation}\label{forces}
\textbf{F}_I = - \frac{\partial E_{DFT+U+V}}{\partial \textbf{R}_I}.
\end{equation}
Formally the forces are computed by means of the Hellman-Feynman theorem. For details on the calculation, we refer to Ref.~\cite{Timrov2020}. Here, we simply note that these forces are then used to update the nuclear positions according to the laws of classical dynamics in the Born-Oppenheimer molecular dynamics.

In summary, we have outlined the essential ingredients of this work. Eqs~\ref{E_uv} and ~\ref{forces} provide the energies and forces necessary for training our equivariant neural network, while Eq~\ref{eq:occup_uv}, the occupation matrix, yields—once diagonalized—the electronic occupations of the Mn $3d$ shells, allowing us to infer its OS.

\subsection*{Computational parameters}

All the calculations were performed using the plane-wave pseudopotential implementation of DFT contained in the \QE distribution~\cite{Giannozzi:2009, Giannozzi:2017}. We use the PBEsol exchange-correlation functional~\cite{Perdew:2008} and pseudopotentials from the SSSP library v1.1~\cite{MaterialsCloud}, which are either ultrasoft (US) or projector-augmented-wave (PAW). For manganese we have used \texttt{mn\_pbesol\_v1.5.uspp.F.UPF} from the GBRV v1.5 library~\cite{Garrity:2014}, for oxygen \texttt{O.pbesol-n-kjpaw\_psl.0.1.UPF} from the Pslibrary v0.3.1~\cite{DalCorso:2014}, for phosphorus \texttt{P.pbesol-n-rrkjus\_psl.1.0.0.UPF} from the Pslibrary v1.0.0~\cite{DalCorso:2014}, and for lithium \texttt{li\_pbesol\_v1.4.uspp.F.UPF} from the GBRV v1.4 library~\cite{Garrity:2014}. All DFT+U+V calculations are performed using orthogonalized atomic orbitals as described in detail in Ref.~\cite{Timrov:2022}.

DFT+U+V FPMD is performed using the $pw.x$ code of \QE. We used averaged self-consistent parameters U=5.1 eV and V=0.7 eV. We use the uniform $\Gamma$-centered \textbf{k}-points grid of size 3×4×5. The KS wavefunctions and potentials are expanded in PWs up to a kinetic-energy cutoff of 65 and 780 Ry, respectively. We used a gaussian smearing with a broadening parameter of 0.005 Ry. We conducted the FPMD simulations on the previous structurally optimized cells with atomic positions relaxed at T=0 K. The Verlet algorithm is used to integrate the classical equations of motion with a timestep of 4 fs. The dynamics was performed in the NVT ensamble (at fixed number of particles N, volume V and temperature T) by using the stochastic velocity rescaling algorithm~\cite{Bussi:2007}. 

For the structural optimizations, the Brillouin zone was sampled using the uniform $\Gamma$-centered \textbf{k} point mesh of size 5×8×9. KS wavefunctions and potentials are expanded in PWs up to a kinetic-energy cutoff of 90 and 1080 Ry, respectively. The crystal structure was optimized using the Broyden-Fletcher-Goldfarb-Shanno (BFGS) algorithm~\cite{Fletcher:1987}, with a convergence threshold for the total energy of $10^{-6}$ Ry, for forces of $10^{-5}$ Ry/Bohr, and for pressure of 0.5 Kbar.

The DFPT calculations of Hubbard parameters are performed using the $hp.x$ code~\cite{Timrov:2022} of \QE using the uniform $\Gamma$-centered \textbf{k} and \textbf{q} point meshes of size 3×4×5 and 1×2×3, respectively. The KS wavefunctions and potentials are expanded in PWs up to a kinetic-energy cutoff of 65 and 780 Ry, respectively, for calculation of Hubbard parameters.

\subsection*{ML architecture}

The model architecture features a convolution filter with a cutoff radius of 4 \AA and consists of four interaction blocks. It employs a maximum rotation order of 2, with feature multiplicities set to 32, including features of odd mirror parity. The radial neural network comprises eight basis radial functions, two radial invariant layers, and 64 hidden neurons. Training was performed over 4200 epochs using the Adam optimizer.


\section*{DATA AVAILABILITY}

The data used to produce the results of this work are available in the Materials Cloud Archive~\cite{malica2024_materialscloud}.


\section*{CODE AVAILABILITY}

DFT+$U$+$V$ calculations are performed using \QE v7.2 which is open-source and can be freely downloaded from https://www.quantum-espresso.org. The machine learning machinery used is NequIP, which is open-source code and can be found https://github.com/mir-group/nequip.


\section*{ACKNOWLEDGEMENTS}

We thank S. Muy and I. Timrov for useful discussions. We gratefully acknowledge support from the Deutsche Forschungsgemeinschaft (DFG) under Germany’s Excellence Strategy (EXC 2077, No. 390741603, University Allowance, University of Bremen) and Lucio Colombi Ciacchi, the host of the “U Bremen Excellence Chair Program”.
We acknowledge support by the European Commission through the MaX Centre of Excellence for supercomputing applications (grant numbers 101093374 and 16HPC096).
We acknowledge support by the NCCR MARVEL, a National Centre of Competence in Research, funded by the Swiss National Science Foundation (grant number 205602).
%


\section*{Author contributions statement}
We use in the following the CRediT (Contributor Roles Taxonomy) author statement.
C.M.: conceptualization, methodology, software, formal analysis, investigation, data curation, writing -- original draft, visualization;
N.M.: supervision, conceptualization, methodology, project administration, funding acquisition. 
All authors: writing -- review \& editing.






\bibliography{main}

\begin{thebibliography}{87}%
\makeatletter
\providecommand \@ifxundefined [1]{%
 \@ifx{#1\undefined}
}%
\providecommand \@ifnum [1]{%
 \ifnum #1\expandafter \@firstoftwo
 \else \expandafter \@secondoftwo
 \fi
}%
\providecommand \@ifx [1]{%
 \ifx #1\expandafter \@firstoftwo
 \else \expandafter \@secondoftwo
 \fi
}%
\providecommand \natexlab [1]{#1}%
\providecommand \enquote  [1]{``#1''}%
\providecommand \bibnamefont  [1]{#1}%
\providecommand \bibfnamefont [1]{#1}%
\providecommand \citenamefont [1]{#1}%
\providecommand \href@noop [0]{\@secondoftwo}%
\providecommand \href [0]{\begingroup \@sanitize@url \@href}%
\providecommand \@href[1]{\@@startlink{#1}\@@href}%
\providecommand \@@href[1]{\endgroup#1\@@endlink}%
\providecommand \@sanitize@url [0]{\catcode `\\12\catcode `\$12\catcode `\&12\catcode `\#12\catcode `\^12\catcode `\_12\catcode `\%12\relax}%
\providecommand \@@startlink[1]{}%
\providecommand \@@endlink[0]{}%
\providecommand \url  [0]{\begingroup\@sanitize@url \@url }%
\providecommand \@url [1]{\endgroup\@href {#1}{\urlprefix }}%
\providecommand \urlprefix  [0]{URL }%
\providecommand \Eprint [0]{\href }%
\providecommand \doibase [0]{https://doi.org/}%
\providecommand \selectlanguage [0]{\@gobble}%
\providecommand \bibinfo  [0]{\@secondoftwo}%
\providecommand \bibfield  [0]{\@secondoftwo}%
\providecommand \translation [1]{[#1]}%
\providecommand \BibitemOpen [0]{}%
\providecommand \bibitemStop [0]{}%
\providecommand \bibitemNoStop [0]{.\EOS\space}%
\providecommand \EOS [0]{\spacefactor3000\relax}%
\providecommand \BibitemShut  [1]{\csname bibitem#1\endcsname}%
\let\auto@bib@innerbib\@empty
\bibitem [{\citenamefont {Walsh}\ \emph {et~al.}(2018)\citenamefont {Walsh}, \citenamefont {Sokol}, \citenamefont {Buckeridge}, \citenamefont {Scanlon},\ and\ \citenamefont {Catlow}}]{Walsh:2018}%
  \BibitemOpen
  \bibfield  {author} {\bibinfo {author} {\bibfnamefont {A.}~\bibnamefont {Walsh}}, \bibinfo {author} {\bibfnamefont {A.~A.}\ \bibnamefont {Sokol}}, \bibinfo {author} {\bibfnamefont {J.}~\bibnamefont {Buckeridge}}, \bibinfo {author} {\bibfnamefont {D.~O.}\ \bibnamefont {Scanlon}},\ and\ \bibinfo {author} {\bibfnamefont {C.~R.~A.}\ \bibnamefont {Catlow}},\ }\bibfield  {title} {\bibinfo {title} {Oxidation states and ionicity},\ }\href {https://doi.org/10.1038/s41563-018-0165-7} {\bibfield  {journal} {\bibinfo  {journal} {Nat. Mater}\ }\textbf {\bibinfo {volume} {17}},\ \bibinfo {pages} {958} (\bibinfo {year} {2018})}\BibitemShut {NoStop}%
\bibitem [{\citenamefont {Barton}(2020)}]{Barton:2020}%
  \BibitemOpen
  \bibfield  {author} {\bibinfo {author} {\bibfnamefont {J.~L.}\ \bibnamefont {Barton}},\ }\bibfield  {title} {\bibinfo {title} {Electrification of the chemical industry},\ }\href {https://doi.org/10.1126/science.abb8061} {\bibfield  {journal} {\bibinfo  {journal} {Science}\ }\textbf {\bibinfo {volume} {368}},\ \bibinfo {pages} {1181} (\bibinfo {year} {2020})},\ \Eprint {https://arxiv.org/abs/https://www.science.org/doi/pdf/10.1126/science.abb8061} {https://www.science.org/doi/pdf/10.1126/science.abb8061} \BibitemShut {NoStop}%
\bibitem [{\citenamefont {Jablonka}\ \emph {et~al.}(2021)\citenamefont {Jablonka}, \citenamefont {Ongari}, \citenamefont {Moosavi},\ and\ \citenamefont {Smit}}]{Jablonka:2021}%
  \BibitemOpen
  \bibfield  {author} {\bibinfo {author} {\bibfnamefont {K.~M.}\ \bibnamefont {Jablonka}}, \bibinfo {author} {\bibfnamefont {D.}~\bibnamefont {Ongari}}, \bibinfo {author} {\bibfnamefont {S.~M.}\ \bibnamefont {Moosavi}},\ and\ \bibinfo {author} {\bibfnamefont {B.}~\bibnamefont {Smit}},\ }\bibfield  {title} {\bibinfo {title} {Using collective knowledge to assign oxidation states of metal cations in metal--organic frameworks},\ }\href {https://doi.org/10.1038/s41557-021-00717-y} {\bibfield  {journal} {\bibinfo  {journal} {Nat. Chem}\ }\textbf {\bibinfo {volume} {13}},\ \bibinfo {pages} {771} (\bibinfo {year} {2021})}\BibitemShut {NoStop}%
\bibitem [{\citenamefont {Nykvist}\ and\ \citenamefont {Nilsson}(2015)}]{Nykvist:2015}%
  \BibitemOpen
  \bibfield  {author} {\bibinfo {author} {\bibfnamefont {B.}~\bibnamefont {Nykvist}}\ and\ \bibinfo {author} {\bibfnamefont {M.}~\bibnamefont {Nilsson}},\ }\bibfield  {title} {\bibinfo {title} {Rapidly falling costs of battery packs for electric vehicles},\ }\href {https://doi.org/10.1038/nclimate2564} {\bibfield  {journal} {\bibinfo  {journal} {Nature Climate Change}\ }\textbf {\bibinfo {volume} {5}},\ \bibinfo {pages} {329} (\bibinfo {year} {2015})}\BibitemShut {NoStop}%
\bibitem [{\citenamefont {Sarma}\ and\ \citenamefont {Shukla}(2018)}]{Sarma:2018}%
  \BibitemOpen
  \bibfield  {author} {\bibinfo {author} {\bibfnamefont {D.~D.}\ \bibnamefont {Sarma}}\ and\ \bibinfo {author} {\bibfnamefont {A.~K.}\ \bibnamefont {Shukla}},\ }\bibfield  {title} {\bibinfo {title} {Building better batteries: A travel back in time},\ }\href {https://doi.org/10.1021/acsenergylett.8b01966} {\bibfield  {journal} {\bibinfo  {journal} {ACS Energy Letters}\ }\textbf {\bibinfo {volume} {3}},\ \bibinfo {pages} {2841} (\bibinfo {year} {2018})}\BibitemShut {NoStop}%
\bibitem [{\citenamefont {Ponrouch}\ and\ \citenamefont {Palacín}(2019)}]{Ponrouch:2019}%
  \BibitemOpen
  \bibfield  {author} {\bibinfo {author} {\bibfnamefont {A.}~\bibnamefont {Ponrouch}}\ and\ \bibinfo {author} {\bibfnamefont {M.~R.}\ \bibnamefont {Palacín}},\ }\bibfield  {title} {\bibinfo {title} {Post-li batteries: promises and challenges},\ }\href {https://doi.org/10.1098/rsta.2018.0297} {\bibfield  {journal} {\bibinfo  {journal} {Phil. Trans. R. Soc. A.}\ }\textbf {\bibinfo {volume} {377}},\ \bibinfo {pages} {37720180297} (\bibinfo {year} {2019})}\BibitemShut {NoStop}%
\bibitem [{\citenamefont {Liu}\ \emph {et~al.}(2019)\citenamefont {Liu}, \citenamefont {Bao}, \citenamefont {Cui}, \citenamefont {Dufek}, \citenamefont {Goodenough}, \citenamefont {Khalifah}, \citenamefont {Li}, \citenamefont {Liaw}, \citenamefont {Liu}, \citenamefont {Manthiram}, \citenamefont {Meng}, \citenamefont {Subramanian}, \citenamefont {Toney}, \citenamefont {Viswanathan}, \citenamefont {Whittingham}, \citenamefont {Xiao}, \citenamefont {Xu}, \citenamefont {Yang}, \citenamefont {Yang},\ and\ \citenamefont {Zhang}}]{Liu:2019}%
  \BibitemOpen
  \bibfield  {author} {\bibinfo {author} {\bibfnamefont {J.}~\bibnamefont {Liu}}, \bibinfo {author} {\bibfnamefont {Z.}~\bibnamefont {Bao}}, \bibinfo {author} {\bibfnamefont {Y.}~\bibnamefont {Cui}}, \bibinfo {author} {\bibfnamefont {E.~J.}\ \bibnamefont {Dufek}}, \bibinfo {author} {\bibfnamefont {J.~B.}\ \bibnamefont {Goodenough}}, \bibinfo {author} {\bibfnamefont {P.}~\bibnamefont {Khalifah}}, \bibinfo {author} {\bibfnamefont {Q.}~\bibnamefont {Li}}, \bibinfo {author} {\bibfnamefont {B.~Y.}\ \bibnamefont {Liaw}}, \bibinfo {author} {\bibfnamefont {P.}~\bibnamefont {Liu}}, \bibinfo {author} {\bibfnamefont {A.}~\bibnamefont {Manthiram}}, \bibinfo {author} {\bibfnamefont {Y.~S.}\ \bibnamefont {Meng}}, \bibinfo {author} {\bibfnamefont {V.~R.}\ \bibnamefont {Subramanian}}, \bibinfo {author} {\bibfnamefont {M.~F.}\ \bibnamefont {Toney}}, \bibinfo {author} {\bibfnamefont {V.~V.}\ \bibnamefont {Viswanathan}}, \bibinfo {author} {\bibfnamefont {M.~S.}\ \bibnamefont {Whittingham}}, \bibinfo {author} {\bibfnamefont
  {J.}~\bibnamefont {Xiao}}, \bibinfo {author} {\bibfnamefont {W.}~\bibnamefont {Xu}}, \bibinfo {author} {\bibfnamefont {J.}~\bibnamefont {Yang}}, \bibinfo {author} {\bibfnamefont {X.-Q.}\ \bibnamefont {Yang}},\ and\ \bibinfo {author} {\bibfnamefont {J.-G.}\ \bibnamefont {Zhang}},\ }\bibfield  {title} {\bibinfo {title} {Pathways for practical high-energy long-cycling lithium metal batteries},\ }\href {https://doi.org/10.1038/s41560-019-0338-x} {\bibfield  {journal} {\bibinfo  {journal} {Nat. Energy}\ }\textbf {\bibinfo {volume} {4}},\ \bibinfo {pages} {180} (\bibinfo {year} {2019})}\BibitemShut {NoStop}%
\bibitem [{\citenamefont {Hohenberg}\ and\ \citenamefont {Kohn}(1964)}]{Hohenberg:1964}%
  \BibitemOpen
  \bibfield  {author} {\bibinfo {author} {\bibfnamefont {P.}~\bibnamefont {Hohenberg}}\ and\ \bibinfo {author} {\bibfnamefont {W.}~\bibnamefont {Kohn}},\ }\bibfield  {title} {\bibinfo {title} {Inhomogeneous electron gas},\ }\href@noop {} {\bibfield  {journal} {\bibinfo  {journal} {Phys. Rev.}\ }\textbf {\bibinfo {volume} {136}},\ \bibinfo {pages} {B864} (\bibinfo {year} {1964})}\BibitemShut {NoStop}%
\bibitem [{\citenamefont {Kohn}\ and\ \citenamefont {Sham}(1965)}]{Kohn:1965}%
  \BibitemOpen
  \bibfield  {author} {\bibinfo {author} {\bibfnamefont {W.}~\bibnamefont {Kohn}}\ and\ \bibinfo {author} {\bibfnamefont {L.}~\bibnamefont {Sham}},\ }\bibfield  {title} {\bibinfo {title} {Self-consistent equations including exchange and correlation effects},\ }\href@noop {} {\bibfield  {journal} {\bibinfo  {journal} {Phys. Rev.}\ }\textbf {\bibinfo {volume} {140}},\ \bibinfo {pages} {A1133} (\bibinfo {year} {1965})}\BibitemShut {NoStop}%
\bibitem [{\citenamefont {Car}\ and\ \citenamefont {Parrinello}(1985)}]{CarParrinello:1985}%
  \BibitemOpen
  \bibfield  {author} {\bibinfo {author} {\bibfnamefont {R.}~\bibnamefont {Car}}\ and\ \bibinfo {author} {\bibfnamefont {M.}~\bibnamefont {Parrinello}},\ }\bibfield  {title} {\bibinfo {title} {Unified approach for molecular dynamics and density-functional theory},\ }\href {https://doi.org/10.1103/PhysRevLett.55.2471} {\bibfield  {journal} {\bibinfo  {journal} {Phys. Rev. Lett.}\ }\textbf {\bibinfo {volume} {55}},\ \bibinfo {pages} {2471} (\bibinfo {year} {1985})}\BibitemShut {NoStop}%
\bibitem [{\citenamefont {McNaught}\ and\ \citenamefont {A.~Wilkinson}(1997)}]{McNaught:1997}%
  \BibitemOpen
  \bibfield  {author} {\bibinfo {author} {\bibfnamefont {A.}~\bibnamefont {McNaught}}\ and\ \bibinfo {author} {\bibfnamefont {I.}~\bibnamefont {A.~Wilkinson}},\ }\href@noop {} {\emph {\bibinfo {title} {Compendium of Chemical Terminology, 2nd ed. (the "Gold Book")}}}\ (\bibinfo  {publisher} {Blackwell Scientific Publications},\ \bibinfo {address} {Oxford},\ \bibinfo {year} {1997})\ \bibinfo {note} {\url{https://doi.org/10.1351/goldbook}}\BibitemShut {NoStop}%
\bibitem [{\citenamefont {Bader}(1990)}]{Bader:1990}%
  \BibitemOpen
  \bibfield  {author} {\bibinfo {author} {\bibfnamefont {R.}~\bibnamefont {Bader}},\ }\href@noop {} {\emph {\bibinfo {title} {{Atoms in Molecules. A Quantum Theory}}}}\ (\bibinfo  {publisher} {Oxford University Press},\ \bibinfo {address} {Oxford, U.K.},\ \bibinfo {year} {1990})\BibitemShut {NoStop}%
\bibitem [{\citenamefont {Bickelhaupt}\ \emph {et~al.}(1996)\citenamefont {Bickelhaupt}, \citenamefont {van Eikema~Hommes}, \citenamefont {Fonseca~Guerra},\ and\ \citenamefont {Baerends}}]{Bickelhaupt:1996}%
  \BibitemOpen
  \bibfield  {author} {\bibinfo {author} {\bibfnamefont {F.~M.}\ \bibnamefont {Bickelhaupt}}, \bibinfo {author} {\bibfnamefont {N.~J.~R.}\ \bibnamefont {van Eikema~Hommes}}, \bibinfo {author} {\bibfnamefont {C.}~\bibnamefont {Fonseca~Guerra}},\ and\ \bibinfo {author} {\bibfnamefont {E.~J.}\ \bibnamefont {Baerends}},\ }\bibfield  {title} {\bibinfo {title} {The carbon-lithium electron pair bond in (ch3li)n (n = 1, 2, 4)},\ }\href {https://doi.org/10.1021/om950966x} {\bibfield  {journal} {\bibinfo  {journal} {Organometallics}\ }\textbf {\bibinfo {volume} {15}},\ \bibinfo {pages} {2923} (\bibinfo {year} {1996})}\BibitemShut {NoStop}%
\bibitem [{\citenamefont {Raebiger}\ \emph {et~al.}(2008)\citenamefont {Raebiger}, \citenamefont {Lany},\ and\ \citenamefont {Zunger}}]{Raebiger:2008}%
  \BibitemOpen
  \bibfield  {author} {\bibinfo {author} {\bibfnamefont {H.}~\bibnamefont {Raebiger}}, \bibinfo {author} {\bibfnamefont {S.}~\bibnamefont {Lany}},\ and\ \bibinfo {author} {\bibfnamefont {A.}~\bibnamefont {Zunger}},\ }\bibfield  {title} {\bibinfo {title} {Charge self-regulation upon changing the oxidation state of transition metals in insulators},\ }\href {https://doi.org/10.1038/nature07009} {\bibfield  {journal} {\bibinfo  {journal} {Nature}\ }\textbf {\bibinfo {volume} {453}},\ \bibinfo {pages} {763} (\bibinfo {year} {2008})}\BibitemShut {NoStop}%
\bibitem [{\citenamefont {Resta}(2008)}]{Resta:2008}%
  \BibitemOpen
  \bibfield  {author} {\bibinfo {author} {\bibfnamefont {R.}~\bibnamefont {Resta}},\ }\bibfield  {title} {\bibinfo {title} {Charge states in transition},\ }\href {https://doi.org/10.1038/453735a} {\bibfield  {journal} {\bibinfo  {journal} {Nature}\ }\textbf {\bibinfo {volume} {453}},\ \bibinfo {pages} {735} (\bibinfo {year} {2008})}\BibitemShut {NoStop}%
\bibitem [{\citenamefont {Jiang}\ \emph {et~al.}(2012)\citenamefont {Jiang}, \citenamefont {Levchenko},\ and\ \citenamefont {Rappe}}]{Rappe:2012}%
  \BibitemOpen
  \bibfield  {author} {\bibinfo {author} {\bibfnamefont {L.}~\bibnamefont {Jiang}}, \bibinfo {author} {\bibfnamefont {S.~V.}\ \bibnamefont {Levchenko}},\ and\ \bibinfo {author} {\bibfnamefont {A.~M.}\ \bibnamefont {Rappe}},\ }\bibfield  {title} {\bibinfo {title} {Rigorous definition of oxidation states of ions in solids},\ }\href {https://doi.org/10.1103/PhysRevLett.108.166403} {\bibfield  {journal} {\bibinfo  {journal} {Phys. Rev. Lett.}\ }\textbf {\bibinfo {volume} {108}},\ \bibinfo {pages} {166403} (\bibinfo {year} {2012})}\BibitemShut {NoStop}%
\bibitem [{\citenamefont {Grasselli}\ and\ \citenamefont {Baroni}(2019)}]{Grasselli:2019}%
  \BibitemOpen
  \bibfield  {author} {\bibinfo {author} {\bibfnamefont {F.}~\bibnamefont {Grasselli}}\ and\ \bibinfo {author} {\bibfnamefont {S.}~\bibnamefont {Baroni}},\ }\bibfield  {title} {\bibinfo {title} {Topological quantization and gauge invariance of charge transport in liquid insulators},\ }\href {https://doi.org/10.1038/s41567-019-0562-0} {\bibfield  {journal} {\bibinfo  {journal} {Nat. Phys}\ }\textbf {\bibinfo {volume} {15}},\ \bibinfo {pages} {967} (\bibinfo {year} {2019})}\BibitemShut {NoStop}%
\bibitem [{\citenamefont {Resta}\ and\ \citenamefont {Vanderbilt}(2007)}]{Resta:2007}%
  \BibitemOpen
  \bibfield  {author} {\bibinfo {author} {\bibfnamefont {R.}~\bibnamefont {Resta}}\ and\ \bibinfo {author} {\bibfnamefont {D.}~\bibnamefont {Vanderbilt}},\ }\bibinfo {title} {Theory of polarization: A modern approach},\ in\ \href {https://doi.org/10.1007/978-3-540-34591-6_2} {\emph {\bibinfo {booktitle} {Physics of Ferroelectrics: A Modern Perspective}}}\ (\bibinfo  {publisher} {Springer Berlin Heidelberg},\ \bibinfo {address} {Berlin, Heidelberg},\ \bibinfo {year} {2007})\ pp.\ \bibinfo {pages} {31--68}\BibitemShut {NoStop}%
\bibitem [{\citenamefont {Vanderbilt}(2018)}]{Vanderbilt:2018}%
  \BibitemOpen
  \bibfield  {author} {\bibinfo {author} {\bibfnamefont {D.}~\bibnamefont {Vanderbilt}},\ }\bibinfo {title} {Berry phases in electronic structure theory: Electric polarization, orbital magnetization and topological insulators}\ (\bibinfo  {publisher} {Cambridge University Press},\ \bibinfo {address} {Cambridge, England},\ \bibinfo {year} {2018})\BibitemShut {NoStop}%
\bibitem [{\citenamefont {Sit}\ \emph {et~al.}(2011)\citenamefont {Sit}, \citenamefont {Car}, \citenamefont {Cohen},\ and\ \citenamefont {Selloni}}]{Sit:2011}%
  \BibitemOpen
  \bibfield  {author} {\bibinfo {author} {\bibfnamefont {P.~H.-L.}\ \bibnamefont {Sit}}, \bibinfo {author} {\bibfnamefont {R.}~\bibnamefont {Car}}, \bibinfo {author} {\bibfnamefont {M.~H.}\ \bibnamefont {Cohen}},\ and\ \bibinfo {author} {\bibfnamefont {A.}~\bibnamefont {Selloni}},\ }\bibfield  {title} {\bibinfo {title} {{Simple, Unambiguous Theoretical Approach to Oxidation State Determination via First-Principles Calculations}},\ }\href@noop {} {\bibfield  {journal} {\bibinfo  {journal} {Inorg. Chem.}\ }\textbf {\bibinfo {volume} {50}},\ \bibinfo {pages} {10259} (\bibinfo {year} {2011})}\BibitemShut {NoStop}%
\bibitem [{\citenamefont {Pegolo}\ \emph {et~al.}(2020)\citenamefont {Pegolo}, \citenamefont {Grasselli},\ and\ \citenamefont {Baroni}}]{Pegolo:2020}%
  \BibitemOpen
  \bibfield  {author} {\bibinfo {author} {\bibfnamefont {P.}~\bibnamefont {Pegolo}}, \bibinfo {author} {\bibfnamefont {F.}~\bibnamefont {Grasselli}},\ and\ \bibinfo {author} {\bibfnamefont {S.}~\bibnamefont {Baroni}},\ }\bibfield  {title} {\bibinfo {title} {{Oxidation States, Thouless’ Pumps, and Nontrivial Ionic Transport in Nonstoichiometric Electrolytes}},\ }\href@noop {} {\bibfield  {journal} {\bibinfo  {journal} {Phys. Rev. X}\ }\textbf {\bibinfo {volume} {10}},\ \bibinfo {pages} {041031} (\bibinfo {year} {2020})}\BibitemShut {NoStop}%
\bibitem [{\citenamefont {Timrov}\ \emph {et~al.}(2022{\natexlab{a}})\citenamefont {Timrov}, \citenamefont {Aquilante}, \citenamefont {Cococcioni},\ and\ \citenamefont {Marzari}}]{Timrov:2022b}%
  \BibitemOpen
  \bibfield  {author} {\bibinfo {author} {\bibfnamefont {I.}~\bibnamefont {Timrov}}, \bibinfo {author} {\bibfnamefont {F.}~\bibnamefont {Aquilante}}, \bibinfo {author} {\bibfnamefont {M.}~\bibnamefont {Cococcioni}},\ and\ \bibinfo {author} {\bibfnamefont {N.}~\bibnamefont {Marzari}},\ }\bibfield  {title} {\bibinfo {title} {Accurate electronic properties and intercalation voltages of olivine-type li-ion cathode materials from extended hubbard functionals},\ }\href {https://doi.org/10.1103/PRXEnergy.1.033003} {\bibfield  {journal} {\bibinfo  {journal} {Phys. Rev. X Energy}\ }\textbf {\bibinfo {volume} {1}},\ \bibinfo {pages} {033003} (\bibinfo {year} {2022}{\natexlab{a}})}\BibitemShut {NoStop}%
\bibitem [{\citenamefont {Hautier}\ \emph {et~al.}(2011{\natexlab{a}})\citenamefont {Hautier}, \citenamefont {Jain}, \citenamefont {Ong}, \citenamefont {Kang}, \citenamefont {Moore}, \citenamefont {Doe},\ and\ \citenamefont {Ceder}}]{Hautier:2011}%
  \BibitemOpen
  \bibfield  {author} {\bibinfo {author} {\bibfnamefont {G.}~\bibnamefont {Hautier}}, \bibinfo {author} {\bibfnamefont {A.}~\bibnamefont {Jain}}, \bibinfo {author} {\bibfnamefont {S.~P.}\ \bibnamefont {Ong}}, \bibinfo {author} {\bibfnamefont {B.}~\bibnamefont {Kang}}, \bibinfo {author} {\bibfnamefont {C.}~\bibnamefont {Moore}}, \bibinfo {author} {\bibfnamefont {R.}~\bibnamefont {Doe}},\ and\ \bibinfo {author} {\bibfnamefont {G.}~\bibnamefont {Ceder}},\ }\bibfield  {title} {\bibinfo {title} {Phosphates as lithium-ion battery cathodes: An evaluation based on high-throughput ab initio calculations},\ }\href {https://doi.org/10.1021/cm200949v} {\bibfield  {journal} {\bibinfo  {journal} {Chem. Mat.}\ }\textbf {\bibinfo {volume} {23}},\ \bibinfo {pages} {3495} (\bibinfo {year} {2011}{\natexlab{a}})}\BibitemShut {NoStop}%
\bibitem [{\citenamefont {Hautier}\ \emph {et~al.}(2011{\natexlab{b}})\citenamefont {Hautier}, \citenamefont {Jain}, \citenamefont {Chen}, \citenamefont {Moore}, \citenamefont {Ong},\ and\ \citenamefont {Ceder}}]{Hautier:2011b}%
  \BibitemOpen
  \bibfield  {author} {\bibinfo {author} {\bibfnamefont {G.}~\bibnamefont {Hautier}}, \bibinfo {author} {\bibfnamefont {A.}~\bibnamefont {Jain}}, \bibinfo {author} {\bibfnamefont {H.}~\bibnamefont {Chen}}, \bibinfo {author} {\bibfnamefont {C.}~\bibnamefont {Moore}}, \bibinfo {author} {\bibfnamefont {S.}~\bibnamefont {Ong}},\ and\ \bibinfo {author} {\bibfnamefont {G.}~\bibnamefont {Ceder}},\ }\bibfield  {title} {\bibinfo {title} {{Novel mixed polyanions lithium-ion battery cathode materials predicted by high-throughput {\it ab initio} computations}},\ }\href@noop {} {\bibfield  {journal} {\bibinfo  {journal} {J. Mater. Chem.}\ }\textbf {\bibinfo {volume} {21}},\ \bibinfo {pages} {17147} (\bibinfo {year} {2011}{\natexlab{b}})}\BibitemShut {NoStop}%
\bibitem [{\citenamefont {Anisimov}\ \emph {et~al.}(1991)\citenamefont {Anisimov}, \citenamefont {Zaanen},\ and\ \citenamefont {Andersen}}]{anisimov:1991}%
  \BibitemOpen
  \bibfield  {author} {\bibinfo {author} {\bibfnamefont {V.}~\bibnamefont {Anisimov}}, \bibinfo {author} {\bibfnamefont {J.}~\bibnamefont {Zaanen}},\ and\ \bibinfo {author} {\bibfnamefont {O.}~\bibnamefont {Andersen}},\ }\bibfield  {title} {\bibinfo {title} {{Band theory and Mott insulators: Hubbard $U$ instead of Stoner $I$}},\ }\href@noop {} {\bibfield  {journal} {\bibinfo  {journal} {Phys. Rev. B}\ }\textbf {\bibinfo {volume} {44}},\ \bibinfo {pages} {943} (\bibinfo {year} {1991})}\BibitemShut {NoStop}%
\bibitem [{\citenamefont {Liechtenstein}\ \emph {et~al.}(1995)\citenamefont {Liechtenstein}, \citenamefont {Anisimov},\ and\ \citenamefont {Zaanen}}]{Liechtenstein:1995}%
  \BibitemOpen
  \bibfield  {author} {\bibinfo {author} {\bibfnamefont {A.}~\bibnamefont {Liechtenstein}}, \bibinfo {author} {\bibfnamefont {V.}~\bibnamefont {Anisimov}},\ and\ \bibinfo {author} {\bibfnamefont {J.}~\bibnamefont {Zaanen}},\ }\bibfield  {title} {\bibinfo {title} {{Density-functional theory and strong interactions: Orbital ordering in Mott-Hubbard insulators}},\ }\href@noop {} {\bibfield  {journal} {\bibinfo  {journal} {Phys. Rev. B}\ }\textbf {\bibinfo {volume} {52}},\ \bibinfo {pages} {R5467} (\bibinfo {year} {1995})}\BibitemShut {NoStop}%
\bibitem [{\citenamefont {Anisimov}\ \emph {et~al.}(1997)\citenamefont {Anisimov}, \citenamefont {Aryasetiawan},\ and\ \citenamefont {Lichtenstein}}]{Anisimov:1997}%
  \BibitemOpen
  \bibfield  {author} {\bibinfo {author} {\bibfnamefont {V.}~\bibnamefont {Anisimov}}, \bibinfo {author} {\bibfnamefont {F.}~\bibnamefont {Aryasetiawan}},\ and\ \bibinfo {author} {\bibfnamefont {A.}~\bibnamefont {Lichtenstein}},\ }\bibfield  {title} {\bibinfo {title} {{First-principles calculations of the electronic structure and spectra of strongly correlated systems: the LDA+$U$ method}},\ }\href@noop {} {\bibfield  {journal} {\bibinfo  {journal} {J. Phys.: Condens. Matter}\ }\textbf {\bibinfo {volume} {9}},\ \bibinfo {pages} {767} (\bibinfo {year} {1997})}\BibitemShut {NoStop}%
\bibitem [{\citenamefont {Dudarev}\ \emph {et~al.}(1998)\citenamefont {Dudarev}, \citenamefont {Botton}, \citenamefont {Savrasov}, \citenamefont {Humphreys},\ and\ \citenamefont {Sutton}}]{Dudarev:1998}%
  \BibitemOpen
  \bibfield  {author} {\bibinfo {author} {\bibfnamefont {S.}~\bibnamefont {Dudarev}}, \bibinfo {author} {\bibfnamefont {G.}~\bibnamefont {Botton}}, \bibinfo {author} {\bibfnamefont {S.}~\bibnamefont {Savrasov}}, \bibinfo {author} {\bibfnamefont {C.}~\bibnamefont {Humphreys}},\ and\ \bibinfo {author} {\bibfnamefont {A.}~\bibnamefont {Sutton}},\ }\bibfield  {title} {\bibinfo {title} {{Electron-energy-loss spectra and the structural stability of nickel oxide: An LSDA+$U$ study}},\ }\href@noop {} {\bibfield  {journal} {\bibinfo  {journal} {Phys. Rev. B}\ }\textbf {\bibinfo {volume} {57}},\ \bibinfo {pages} {1505} (\bibinfo {year} {1998})}\BibitemShut {NoStop}%
\bibitem [{\citenamefont {Campo~Jr}\ and\ \citenamefont {Cococcioni}(2010)}]{Campo:2010}%
  \BibitemOpen
  \bibfield  {author} {\bibinfo {author} {\bibfnamefont {V.~L.}\ \bibnamefont {Campo~Jr}}\ and\ \bibinfo {author} {\bibfnamefont {M.}~\bibnamefont {Cococcioni}},\ }\bibfield  {title} {\bibinfo {title} {{Extended DFT+$U$+$V$ method with on-site and inter-site electronic interactions}},\ }\href@noop {} {\bibfield  {journal} {\bibinfo  {journal} {J. Phys.: Condens. Matter}\ }\textbf {\bibinfo {volume} {22}},\ \bibinfo {pages} {055602} (\bibinfo {year} {2010})}\BibitemShut {NoStop}%
\bibitem [{\citenamefont {Tancogne-Dejean}\ and\ \citenamefont {Rubio}(2020)}]{TancogneDejean:2020}%
  \BibitemOpen
  \bibfield  {author} {\bibinfo {author} {\bibfnamefont {N.}~\bibnamefont {Tancogne-Dejean}}\ and\ \bibinfo {author} {\bibfnamefont {A.}~\bibnamefont {Rubio}},\ }\bibfield  {title} {\bibinfo {title} {{Parameter-free hybridlike functional based on an extended Hubbard model: DFT+U+V}},\ }\href@noop {} {\bibfield  {journal} {\bibinfo  {journal} {Phys. Rev. B}\ }\textbf {\bibinfo {volume} {102}},\ \bibinfo {pages} {155117} (\bibinfo {year} {2020})}\BibitemShut {NoStop}%
\bibitem [{\citenamefont {Lee}\ and\ \citenamefont {Son}(2020)}]{Lee:2020}%
  \BibitemOpen
  \bibfield  {author} {\bibinfo {author} {\bibfnamefont {S.-H.}\ \bibnamefont {Lee}}\ and\ \bibinfo {author} {\bibfnamefont {Y.-W.}\ \bibnamefont {Son}},\ }\bibfield  {title} {\bibinfo {title} {{First-principles approach with a pseudohybrid density functional for extended Hubbard interactions}},\ }\href@noop {} {\bibfield  {journal} {\bibinfo  {journal} {Phys. Rev. Res.}\ }\textbf {\bibinfo {volume} {2}},\ \bibinfo {pages} {043410} (\bibinfo {year} {2020})}\BibitemShut {NoStop}%
\bibitem [{\citenamefont {Sun}\ \emph {et~al.}(2015)\citenamefont {Sun}, \citenamefont {Ruzsinszky},\ and\ \citenamefont {Perdew}}]{Sun:2015}%
  \BibitemOpen
  \bibfield  {author} {\bibinfo {author} {\bibfnamefont {J.}~\bibnamefont {Sun}}, \bibinfo {author} {\bibfnamefont {A.}~\bibnamefont {Ruzsinszky}},\ and\ \bibinfo {author} {\bibfnamefont {J.}~\bibnamefont {Perdew}},\ }\bibfield  {title} {\bibinfo {title} {{Strongly Constrained and Appropriately Normed Semilocal Density Functional}},\ }\href@noop {} {\bibfield  {journal} {\bibinfo  {journal} {Phys. Rev. Lett.}\ }\textbf {\bibinfo {volume} {115}},\ \bibinfo {pages} {036402} (\bibinfo {year} {2015})}\BibitemShut {NoStop}%
\bibitem [{\citenamefont {Bart\'ok}\ and\ \citenamefont {Yates}(2019)}]{Bartok:2019}%
  \BibitemOpen
  \bibfield  {author} {\bibinfo {author} {\bibfnamefont {A.}~\bibnamefont {Bart\'ok}}\ and\ \bibinfo {author} {\bibfnamefont {J.}~\bibnamefont {Yates}},\ }\bibfield  {title} {\bibinfo {title} {{Regularized SCAN functional}},\ }\href@noop {} {\bibfield  {journal} {\bibinfo  {journal} {J. Chem. Phys.}\ }\textbf {\bibinfo {volume} {150}},\ \bibinfo {pages} {161101} (\bibinfo {year} {2019})}\BibitemShut {NoStop}%
\bibitem [{\citenamefont {Furness}\ \emph {et~al.}(2020)\citenamefont {Furness}, \citenamefont {Kaplan}, \citenamefont {Ning}, \citenamefont {Perdew},\ and\ \citenamefont {Sun}}]{Furness:2020}%
  \BibitemOpen
  \bibfield  {author} {\bibinfo {author} {\bibfnamefont {J.}~\bibnamefont {Furness}}, \bibinfo {author} {\bibfnamefont {A.}~\bibnamefont {Kaplan}}, \bibinfo {author} {\bibfnamefont {J.}~\bibnamefont {Ning}}, \bibinfo {author} {\bibfnamefont {J.}~\bibnamefont {Perdew}},\ and\ \bibinfo {author} {\bibfnamefont {J.}~\bibnamefont {Sun}},\ }\bibfield  {title} {\bibinfo {title} {{Accurate and Numerically Efficient r$^2$SCAN Meta-Generalized Gradient Approximation}},\ }\href@noop {} {\bibfield  {journal} {\bibinfo  {journal} {J. Phys. Chem. Lett.}\ }\textbf {\bibinfo {volume} {11}},\ \bibinfo {pages} {8208} (\bibinfo {year} {2020})}\BibitemShut {NoStop}%
\bibitem [{\citenamefont {Gautam}\ and\ \citenamefont {Carter}(2018)}]{Gautam:2018}%
  \BibitemOpen
  \bibfield  {author} {\bibinfo {author} {\bibfnamefont {G.}~\bibnamefont {Gautam}}\ and\ \bibinfo {author} {\bibfnamefont {E.}~\bibnamefont {Carter}},\ }\bibfield  {title} {\bibinfo {title} {{Evaluating transition metal oxides within DFT-SCAN and SCAN+U frameworks for solar thermochemical applications}},\ }\href@noop {} {\bibfield  {journal} {\bibinfo  {journal} {Phys. Rev. Mater.}\ }\textbf {\bibinfo {volume} {2}},\ \bibinfo {pages} {095401} (\bibinfo {year} {2018})}\BibitemShut {NoStop}%
\bibitem [{\citenamefont {Long}\ \emph {et~al.}(2020)\citenamefont {Long}, \citenamefont {Gautam},\ and\ \citenamefont {Carter}}]{Long:2020}%
  \BibitemOpen
  \bibfield  {author} {\bibinfo {author} {\bibfnamefont {O.}~\bibnamefont {Long}}, \bibinfo {author} {\bibfnamefont {G.}~\bibnamefont {Gautam}},\ and\ \bibinfo {author} {\bibfnamefont {E.}~\bibnamefont {Carter}},\ }\bibfield  {title} {\bibinfo {title} {{Evaluating optimal $U$ for $3d$ transition-metal oxides within the SCAN+$U$ framework}},\ }\href@noop {} {\bibfield  {journal} {\bibinfo  {journal} {Phys. Rev. Mater.}\ }\textbf {\bibinfo {volume} {4}},\ \bibinfo {pages} {045401} (\bibinfo {year} {2020})}\BibitemShut {NoStop}%
\bibitem [{\citenamefont {Kaczkowski}\ \emph {et~al.}(2021)\citenamefont {Kaczkowski}, \citenamefont {Pugaczowa-Michalska},\ and\ \citenamefont {P\'lowa\'s-Korus}}]{Kaczkowski:2021}%
  \BibitemOpen
  \bibfield  {author} {\bibinfo {author} {\bibfnamefont {J.}~\bibnamefont {Kaczkowski}}, \bibinfo {author} {\bibfnamefont {M.}~\bibnamefont {Pugaczowa-Michalska}},\ and\ \bibinfo {author} {\bibfnamefont {I.}~\bibnamefont {P\'lowa\'s-Korus}},\ }\bibfield  {title} {\bibinfo {title} {{Comparative density functional studies of pristine and doped bismuth ferrite polymorphs by GGA+$U$ and meta-GGA SCAN+$U$}},\ }\href@noop {} {\bibfield  {journal} {\bibinfo  {journal} {Phys. Chem. Chem. Phys.}\ }\textbf {\bibinfo {volume} {23}},\ \bibinfo {pages} {8571} (\bibinfo {year} {2021})}\BibitemShut {NoStop}%
\bibitem [{\citenamefont {Artrith}\ \emph {et~al.}(2022)\citenamefont {Artrith}, \citenamefont {Torres}, \citenamefont {Urban},\ and\ \citenamefont {Hybertsen}}]{Artrith:2022}%
  \BibitemOpen
  \bibfield  {author} {\bibinfo {author} {\bibfnamefont {N.}~\bibnamefont {Artrith}}, \bibinfo {author} {\bibfnamefont {J.}~\bibnamefont {Torres}}, \bibinfo {author} {\bibfnamefont {A.}~\bibnamefont {Urban}},\ and\ \bibinfo {author} {\bibfnamefont {M.}~\bibnamefont {Hybertsen}},\ }\bibfield  {title} {\bibinfo {title} {{Data-driven approach to parameterize SCAN+$U$ for an accurate description of $3d$ transition metal oxide thermochemistry}},\ }\href@noop {} {\bibfield  {journal} {\bibinfo  {journal} {Phys. Rev. Mater.}\ }\textbf {\bibinfo {volume} {6}},\ \bibinfo {pages} {035003} (\bibinfo {year} {2022})}\BibitemShut {NoStop}%
\bibitem [{\citenamefont {Adamo}\ and\ \citenamefont {Barone}(1999)}]{Adamo:1999}%
  \BibitemOpen
  \bibfield  {author} {\bibinfo {author} {\bibfnamefont {C.}~\bibnamefont {Adamo}}\ and\ \bibinfo {author} {\bibfnamefont {V.}~\bibnamefont {Barone}},\ }\bibfield  {title} {\bibinfo {title} {{Toward reliable density functional methods without adjustable parameters: The PBE0 model}},\ }\href@noop {} {\bibfield  {journal} {\bibinfo  {journal} {J. Chem. Phys.}\ }\textbf {\bibinfo {volume} {110}},\ \bibinfo {pages} {6158} (\bibinfo {year} {1999})}\BibitemShut {NoStop}%
\bibitem [{\citenamefont {Heyd}\ \emph {et~al.}(2003)\citenamefont {Heyd}, \citenamefont {Scuseria},\ and\ \citenamefont {Ernzerhof}}]{Heyd:2003}%
  \BibitemOpen
  \bibfield  {author} {\bibinfo {author} {\bibfnamefont {J.}~\bibnamefont {Heyd}}, \bibinfo {author} {\bibfnamefont {G.}~\bibnamefont {Scuseria}},\ and\ \bibinfo {author} {\bibfnamefont {M.}~\bibnamefont {Ernzerhof}},\ }\bibfield  {title} {\bibinfo {title} {{Hybrid functionals based on a screened Coulomb potential}},\ }\href@noop {} {\bibfield  {journal} {\bibinfo  {journal} {J. Chem. Phys.}\ }\textbf {\bibinfo {volume} {118}},\ \bibinfo {pages} {8207} (\bibinfo {year} {2003})}\BibitemShut {NoStop}%
\bibitem [{\citenamefont {Heyd}\ \emph {et~al.}(2006)\citenamefont {Heyd}, \citenamefont {Scuseria},\ and\ \citenamefont {Ernzerhof}}]{Heyd:2006}%
  \BibitemOpen
  \bibfield  {author} {\bibinfo {author} {\bibfnamefont {J.}~\bibnamefont {Heyd}}, \bibinfo {author} {\bibfnamefont {G.}~\bibnamefont {Scuseria}},\ and\ \bibinfo {author} {\bibfnamefont {M.}~\bibnamefont {Ernzerhof}},\ }\bibfield  {title} {\bibinfo {title} {{Erratum: “Hybrid functionals based on a screened Coulomb potential” [J. Chem. Phys. 118, 8207 (2003)]}},\ }\href@noop {} {\bibfield  {journal} {\bibinfo  {journal} {J. Chem. Phys.}\ }\textbf {\bibinfo {volume} {124}},\ \bibinfo {pages} {219906} (\bibinfo {year} {2006})}\BibitemShut {NoStop}%
\bibitem [{\citenamefont {Cococcioni}\ and\ \citenamefont {Marzari}(2019)}]{Cococcioni:2019}%
  \BibitemOpen
  \bibfield  {author} {\bibinfo {author} {\bibfnamefont {M.}~\bibnamefont {Cococcioni}}\ and\ \bibinfo {author} {\bibfnamefont {N.}~\bibnamefont {Marzari}},\ }\bibfield  {title} {\bibinfo {title} {{Energetics and cathode voltages of Li$M$PO$_4$ olivines ($M$=Fe, Mn) from extended Hubbard functionals}},\ }\href@noop {} {\bibfield  {journal} {\bibinfo  {journal} {Phys. Rev. Materials}\ }\textbf {\bibinfo {volume} {3}},\ \bibinfo {pages} {033801} (\bibinfo {year} {2019})}\BibitemShut {NoStop}%
\bibitem [{\citenamefont {Sit}\ \emph {et~al.}(2006)\citenamefont {Sit}, \citenamefont {Cococcioni},\ and\ \citenamefont {Marzari}}]{Sit:2006}%
  \BibitemOpen
  \bibfield  {author} {\bibinfo {author} {\bibfnamefont {P.~H.-L.}\ \bibnamefont {Sit}}, \bibinfo {author} {\bibfnamefont {M.}~\bibnamefont {Cococcioni}},\ and\ \bibinfo {author} {\bibfnamefont {N.}~\bibnamefont {Marzari}},\ }\bibfield  {title} {\bibinfo {title} {Realistic quantitative descriptions of electron transfer reactions: Diabatic free-energy surfaces from first-principles molecular dynamics},\ }\href {https://doi.org/10.1103/PhysRevLett.97.028303} {\bibfield  {journal} {\bibinfo  {journal} {Phys. Rev. Lett.}\ }\textbf {\bibinfo {volume} {97}},\ \bibinfo {pages} {028303} (\bibinfo {year} {2006})}\BibitemShut {NoStop}%
\bibitem [{\citenamefont {Newton}\ and\ \citenamefont {Sutin}(1984)}]{Newton:1984}%
  \BibitemOpen
  \bibfield  {author} {\bibinfo {author} {\bibfnamefont {M.~D.}\ \bibnamefont {Newton}}\ and\ \bibinfo {author} {\bibfnamefont {N.}~\bibnamefont {Sutin}},\ }\bibfield  {title} {\bibinfo {title} {Electron transfer reactions in condensed phases},\ }\href@noop {} {\bibfield  {journal} {\bibinfo  {journal} {Annu. Rev. Phys. Chem.}\ }\textbf {\bibinfo {volume} {35}},\ \bibinfo {pages} {437} (\bibinfo {year} {1984})}\BibitemShut {NoStop}%
\bibitem [{\citenamefont {Marcus}\ and\ \citenamefont {Sutin}(1985)}]{Marcus:1985}%
  \BibitemOpen
  \bibfield  {author} {\bibinfo {author} {\bibfnamefont {R.}~\bibnamefont {Marcus}}\ and\ \bibinfo {author} {\bibfnamefont {N.}~\bibnamefont {Sutin}},\ }\bibfield  {title} {\bibinfo {title} {Electron transfers in chemistry and biology},\ }\href@noop {} {\bibfield  {journal} {\bibinfo  {journal} {Biochimica et Biophysica Acta (BBA) - Reviews on Bioenergetics}\ }\textbf {\bibinfo {volume} {811}},\ \bibinfo {pages} {265} (\bibinfo {year} {1985})}\BibitemShut {NoStop}%
\bibitem [{\citenamefont {Warshel}\ and\ \citenamefont {Parson}(1991)}]{Warshel:1991}%
  \BibitemOpen
  \bibfield  {author} {\bibinfo {author} {\bibfnamefont {A.}~\bibnamefont {Warshel}}\ and\ \bibinfo {author} {\bibfnamefont {W.~W.}\ \bibnamefont {Parson}},\ }\bibfield  {title} {\bibinfo {title} {Computer simulations of electron-transfer reactions in solution and in photosynthetic reaction centers},\ }\href@noop {} {\bibfield  {journal} {\bibinfo  {journal} {Annu. Rev. Phys. Chem.}\ }\textbf {\bibinfo {volume} {42}},\ \bibinfo {pages} {279} (\bibinfo {year} {1991})}\BibitemShut {NoStop}%
\bibitem [{\citenamefont {Behler}\ and\ \citenamefont {Parrinello}(2007)}]{Behler:2007}%
  \BibitemOpen
  \bibfield  {author} {\bibinfo {author} {\bibfnamefont {J.}~\bibnamefont {Behler}}\ and\ \bibinfo {author} {\bibfnamefont {M.}~\bibnamefont {Parrinello}},\ }\bibfield  {title} {\bibinfo {title} {Generalized neural-network representation of high-dimensional potential-energy surfaces},\ }\href {https://doi.org/10.1103/PhysRevLett.98.146401} {\bibfield  {journal} {\bibinfo  {journal} {Phys. Rev. Lett.}\ }\textbf {\bibinfo {volume} {98}},\ \bibinfo {pages} {146401} (\bibinfo {year} {2007})}\BibitemShut {NoStop}%
\bibitem [{\citenamefont {Batzner}\ \emph {et~al.}(2022)\citenamefont {Batzner}, \citenamefont {Musaelian}, \citenamefont {Sun}, \citenamefont {Geiger}, \citenamefont {Mailoa}, \citenamefont {Kornbluth}, \citenamefont {Molinari}, \citenamefont {Smidt},\ and\ \citenamefont {Kozinsky}}]{Batzner:2022}%
  \BibitemOpen
  \bibfield  {author} {\bibinfo {author} {\bibfnamefont {S.}~\bibnamefont {Batzner}}, \bibinfo {author} {\bibfnamefont {A.}~\bibnamefont {Musaelian}}, \bibinfo {author} {\bibfnamefont {L.}~\bibnamefont {Sun}}, \bibinfo {author} {\bibfnamefont {M.}~\bibnamefont {Geiger}}, \bibinfo {author} {\bibfnamefont {J.~P.}\ \bibnamefont {Mailoa}}, \bibinfo {author} {\bibfnamefont {M.}~\bibnamefont {Kornbluth}}, \bibinfo {author} {\bibfnamefont {N.}~\bibnamefont {Molinari}}, \bibinfo {author} {\bibfnamefont {T.~E.}\ \bibnamefont {Smidt}},\ and\ \bibinfo {author} {\bibfnamefont {B.}~\bibnamefont {Kozinsky}},\ }\bibfield  {title} {\bibinfo {title} {E(3)-equivariant graph neural networks for data-efficient and accurate interatomic potentials},\ }\href@noop {} {\bibfield  {journal} {\bibinfo  {journal} {Nat. Commun.}\ }\textbf {\bibinfo {volume} {13}},\ \bibinfo {pages} {2453} (\bibinfo {year} {2022})}\BibitemShut {NoStop}%
\bibitem [{\citenamefont {Batatia}\ \emph {et~al.}(2022)\citenamefont {Batatia}, \citenamefont {Kovacs}, \citenamefont {Simm}, \citenamefont {Ortner},\ and\ \citenamefont {Csanyi}}]{Batatia:2022}%
  \BibitemOpen
  \bibfield  {author} {\bibinfo {author} {\bibfnamefont {I.}~\bibnamefont {Batatia}}, \bibinfo {author} {\bibfnamefont {D.~P.}\ \bibnamefont {Kovacs}}, \bibinfo {author} {\bibfnamefont {G.}~\bibnamefont {Simm}}, \bibinfo {author} {\bibfnamefont {C.}~\bibnamefont {Ortner}},\ and\ \bibinfo {author} {\bibfnamefont {G.}~\bibnamefont {Csanyi}},\ }\bibfield  {title} {\bibinfo {title} {Mace: Higher order equivariant message passing neural networks for fast and accurate force fields},\ }in\ \href {https://proceedings.neurips.cc/paper_files/paper/2022/file/4a36c3c51af11ed9f34615b81edb5bbc-Paper-Conference.pdf} {\emph {\bibinfo {booktitle} {Advances in Neural Information Processing Systems}}},\ Vol.~\bibinfo {volume} {35},\ \bibinfo {editor} {edited by\ \bibinfo {editor} {\bibfnamefont {S.}~\bibnamefont {Koyejo}}, \bibinfo {editor} {\bibfnamefont {S.}~\bibnamefont {Mohamed}}, \bibinfo {editor} {\bibfnamefont {A.}~\bibnamefont {Agarwal}}, \bibinfo {editor} {\bibfnamefont {D.}~\bibnamefont {Belgrave}}, \bibinfo {editor}
  {\bibfnamefont {K.}~\bibnamefont {Cho}},\ and\ \bibinfo {editor} {\bibfnamefont {A.}~\bibnamefont {Oh}}}\ (\bibinfo  {publisher} {Curran Associates, Inc.},\ \bibinfo {year} {2022})\ pp.\ \bibinfo {pages} {11423--11436}\BibitemShut {NoStop}%
\bibitem [{\citenamefont {Zhang}\ \emph {et~al.}(2018)\citenamefont {Zhang}, \citenamefont {Han}, \citenamefont {Wang}, \citenamefont {Car},\ and\ \citenamefont {E}}]{Zhang:2018}%
  \BibitemOpen
  \bibfield  {author} {\bibinfo {author} {\bibfnamefont {L.}~\bibnamefont {Zhang}}, \bibinfo {author} {\bibfnamefont {J.}~\bibnamefont {Han}}, \bibinfo {author} {\bibfnamefont {H.}~\bibnamefont {Wang}}, \bibinfo {author} {\bibfnamefont {R.}~\bibnamefont {Car}},\ and\ \bibinfo {author} {\bibfnamefont {W.}~\bibnamefont {E}},\ }\bibfield  {title} {\bibinfo {title} {Deep potential molecular dynamics: A scalable model with the accuracy of quantum mechanics},\ }\href {https://doi.org/10.1103/PhysRevLett.120.143001} {\bibfield  {journal} {\bibinfo  {journal} {Phys. Rev. Lett.}\ }\textbf {\bibinfo {volume} {120}},\ \bibinfo {pages} {143001} (\bibinfo {year} {2018})}\BibitemShut {NoStop}%
\bibitem [{\citenamefont {Reed}\ and\ \citenamefont {Ceder}(2004)}]{Reed:2004}%
  \BibitemOpen
  \bibfield  {author} {\bibinfo {author} {\bibfnamefont {J.}~\bibnamefont {Reed}}\ and\ \bibinfo {author} {\bibfnamefont {G.}~\bibnamefont {Ceder}},\ }\bibfield  {title} {\bibinfo {title} {Role of electronic structure in the susceptibility of metastable transition-metal oxide structures to transformation},\ }\href {https://doi.org/10.1021/cr020733x} {\bibfield  {journal} {\bibinfo  {journal} {Chem. Rev.}\ }\textbf {\bibinfo {volume} {104}},\ \bibinfo {pages} {4513} (\bibinfo {year} {2004})}\BibitemShut {NoStop}%
\bibitem [{\citenamefont {Rappe}\ and\ \citenamefont {Goddard~III}(1991)}]{Rappe:1991}%
  \BibitemOpen
  \bibfield  {author} {\bibinfo {author} {\bibfnamefont {A.~K.}\ \bibnamefont {Rappe}}\ and\ \bibinfo {author} {\bibfnamefont {W.~A.}\ \bibnamefont {Goddard~III}},\ }\bibfield  {title} {\bibinfo {title} {Charge equilibration for molecular dynamics simulations},\ }\href {https://doi.org/10.1021/j100161a070} {\bibfield  {journal} {\bibinfo  {journal} {The Journal of Physical Chemistry}\ }\textbf {\bibinfo {volume} {95}},\ \bibinfo {pages} {3358} (\bibinfo {year} {1991})}\BibitemShut {NoStop}%
\bibitem [{\citenamefont {Ghasemi}\ \emph {et~al.}(2015)\citenamefont {Ghasemi}, \citenamefont {Hofstetter}, \citenamefont {Saha},\ and\ \citenamefont {Goedecker}}]{Ghasemi:2015}%
  \BibitemOpen
  \bibfield  {author} {\bibinfo {author} {\bibfnamefont {S.~A.}\ \bibnamefont {Ghasemi}}, \bibinfo {author} {\bibfnamefont {A.}~\bibnamefont {Hofstetter}}, \bibinfo {author} {\bibfnamefont {S.}~\bibnamefont {Saha}},\ and\ \bibinfo {author} {\bibfnamefont {S.}~\bibnamefont {Goedecker}},\ }\bibfield  {title} {\bibinfo {title} {Interatomic potentials for ionic systems with density functional accuracy based on charge densities obtained by a neural network},\ }\href {https://doi.org/10.1103/PhysRevB.92.045131} {\bibfield  {journal} {\bibinfo  {journal} {Phys. Rev. B}\ }\textbf {\bibinfo {volume} {92}},\ \bibinfo {pages} {045131} (\bibinfo {year} {2015})}\BibitemShut {NoStop}%
\bibitem [{\citenamefont {Ko}\ \emph {et~al.}(2021)\citenamefont {Ko}, \citenamefont {Finkler}, \citenamefont {Goedecker},\ and\ \citenamefont {Behler}}]{Ko:2021}%
  \BibitemOpen
  \bibfield  {author} {\bibinfo {author} {\bibfnamefont {T.~W.}\ \bibnamefont {Ko}}, \bibinfo {author} {\bibfnamefont {J.~A.}\ \bibnamefont {Finkler}}, \bibinfo {author} {\bibfnamefont {S.}~\bibnamefont {Goedecker}},\ and\ \bibinfo {author} {\bibfnamefont {J.}~\bibnamefont {Behler}},\ }\bibfield  {title} {\bibinfo {title} {A fourth-generation high-dimensional neural network potential with accurate electrostatics including non-local charge transfer},\ }\href@noop {} {\bibfield  {journal} {\bibinfo  {journal} {Nat Commun}\ }\textbf {\bibinfo {volume} {12}},\ \bibinfo {pages} {398} (\bibinfo {year} {2021})}\BibitemShut {NoStop}%
\bibitem [{\citenamefont {Staacke}\ \emph {et~al.}(2022)\citenamefont {Staacke}, \citenamefont {Wengert}, \citenamefont {Kunkel}, \citenamefont {Csányi}, \citenamefont {Reuter},\ and\ \citenamefont {Margraf}}]{Staacke:2022}%
  \BibitemOpen
  \bibfield  {author} {\bibinfo {author} {\bibfnamefont {C.~G.}\ \bibnamefont {Staacke}}, \bibinfo {author} {\bibfnamefont {S.}~\bibnamefont {Wengert}}, \bibinfo {author} {\bibfnamefont {C.}~\bibnamefont {Kunkel}}, \bibinfo {author} {\bibfnamefont {G.}~\bibnamefont {Csányi}}, \bibinfo {author} {\bibfnamefont {K.}~\bibnamefont {Reuter}},\ and\ \bibinfo {author} {\bibfnamefont {J.~T.}\ \bibnamefont {Margraf}},\ }\bibfield  {title} {\bibinfo {title} {Kernel charge equilibration: efficient and accurate prediction of molecular dipole moments with a machine-learning enhanced electron density model},\ }\href {https://doi.org/10.1088/2632-2153/ac568d} {\bibfield  {journal} {\bibinfo  {journal} {Mach. Learn.: Sci. Technol.}\ }\textbf {\bibinfo {volume} {3}},\ \bibinfo {pages} {015032} (\bibinfo {year} {2022})}\BibitemShut {NoStop}%
\bibitem [{\citenamefont {Shaidu}\ \emph {et~al.}(2024)\citenamefont {Shaidu}, \citenamefont {Pellegrini}, \citenamefont {K{\"u}{\c{c}}{\"u}kbenli}, \citenamefont {Lot},\ and\ \citenamefont {de~Gironcoli}}]{Shaidu:2024}%
  \BibitemOpen
  \bibfield  {author} {\bibinfo {author} {\bibfnamefont {Y.}~\bibnamefont {Shaidu}}, \bibinfo {author} {\bibfnamefont {F.}~\bibnamefont {Pellegrini}}, \bibinfo {author} {\bibfnamefont {E.}~\bibnamefont {K{\"u}{\c{c}}{\"u}kbenli}}, \bibinfo {author} {\bibfnamefont {R.}~\bibnamefont {Lot}},\ and\ \bibinfo {author} {\bibfnamefont {S.}~\bibnamefont {de~Gironcoli}},\ }\bibfield  {title} {\bibinfo {title} {Incorporating long-range electrostatics in neural network potentials via variational charge equilibration from shortsighted ingredients},\ }\href {https://doi.org/10.1038/s41524-024-01225-6} {\bibfield  {journal} {\bibinfo  {journal} {Npj Comput. Mater.}\ }\textbf {\bibinfo {volume} {10}},\ \bibinfo {pages} {47} (\bibinfo {year} {2024})}\BibitemShut {NoStop}%
\bibitem [{\citenamefont {Chen}\ \emph {et~al.}(2023)\citenamefont {Chen}, \citenamefont {El~Khatib}, \citenamefont {Lindgren}, \citenamefont {Willard}, \citenamefont {Medford},\ and\ \citenamefont {Peterson}}]{Chen:2023}%
  \BibitemOpen
  \bibfield  {author} {\bibinfo {author} {\bibfnamefont {X.}~\bibnamefont {Chen}}, \bibinfo {author} {\bibfnamefont {M.}~\bibnamefont {El~Khatib}}, \bibinfo {author} {\bibfnamefont {P.}~\bibnamefont {Lindgren}}, \bibinfo {author} {\bibfnamefont {A.}~\bibnamefont {Willard}}, \bibinfo {author} {\bibfnamefont {A.~J.}\ \bibnamefont {Medford}},\ and\ \bibinfo {author} {\bibfnamefont {A.~A.}\ \bibnamefont {Peterson}},\ }\bibfield  {title} {\bibinfo {title} {Atomistic learning in the electronically grand-canonical ensemble},\ }\href@noop {} {\bibfield  {journal} {\bibinfo  {journal} {Npj Comput. Mater.}\ }\textbf {\bibinfo {volume} {9}} (\bibinfo {year} {2023})}\BibitemShut {NoStop}%
\bibitem [{\citenamefont {Kocer}\ \emph {et~al.}(2024)\citenamefont {Kocer}, \citenamefont {Haouari}, \citenamefont {Dellago},\ and\ \citenamefont {Behler}}]{kocer:2024}%
  \BibitemOpen
  \bibfield  {author} {\bibinfo {author} {\bibfnamefont {E.}~\bibnamefont {Kocer}}, \bibinfo {author} {\bibfnamefont {R.~E.}\ \bibnamefont {Haouari}}, \bibinfo {author} {\bibfnamefont {C.}~\bibnamefont {Dellago}},\ and\ \bibinfo {author} {\bibfnamefont {J.}~\bibnamefont {Behler}},\ }\href {https://arxiv.org/abs/2410.03299} {\bibinfo {title} {Machine learning potentials for redox chemistry in solution}} (\bibinfo {year} {2024}),\ \Eprint {https://arxiv.org/abs/2410.03299} {arXiv:2410.03299 [physics.chem-ph]} \BibitemShut {NoStop}%
\bibitem [{\citenamefont {Deng}\ \emph {et~al.}(2023)\citenamefont {Deng}, \citenamefont {Ahong}, \citenamefont {Jun}, \citenamefont {Riebesell},\ and\ \citenamefont {Han}}]{Deng:2023}%
  \BibitemOpen
  \bibfield  {author} {\bibinfo {author} {\bibfnamefont {B.}~\bibnamefont {Deng}}, \bibinfo {author} {\bibfnamefont {P.}~\bibnamefont {Ahong}}, \bibinfo {author} {\bibfnamefont {K.}~\bibnamefont {Jun}}, \bibinfo {author} {\bibfnamefont {J.}~\bibnamefont {Riebesell}},\ and\ \bibinfo {author} {\bibfnamefont {K.}~\bibnamefont {Han}},\ }\bibfield  {title} {\bibinfo {title} {Chgnet as a pretrained universal neural network potential for charge-informed atomistic modelling},\ }\href@noop {} {\bibfield  {journal} {\bibinfo  {journal} {Nat. Mach. Intell.}\ }\textbf {\bibinfo {volume} {5}} (\bibinfo {year} {2023})}\BibitemShut {NoStop}%
\bibitem [{\citenamefont {Eckhoff}\ \emph {et~al.}(2020)\citenamefont {Eckhoff}, \citenamefont {Lausch}, \citenamefont {Bl{\"o}chl},\ and\ \citenamefont {Behler}}]{Eckhoff:2020}%
  \BibitemOpen
  \bibfield  {author} {\bibinfo {author} {\bibfnamefont {M.}~\bibnamefont {Eckhoff}}, \bibinfo {author} {\bibfnamefont {K.~N.}\ \bibnamefont {Lausch}}, \bibinfo {author} {\bibfnamefont {P.~E.}\ \bibnamefont {Bl{\"o}chl}},\ and\ \bibinfo {author} {\bibfnamefont {J.}~\bibnamefont {Behler}},\ }\bibfield  {title} {\bibinfo {title} {Predicting oxidation and spin states by high-dimensional neural networks: Applications to lithium manganese oxide spinels},\ }\href@noop {} {\bibfield  {journal} {\bibinfo  {journal} {J. Chem. Phys.}\ }\textbf {\bibinfo {volume} {153}},\ \bibinfo {pages} {164107} (\bibinfo {year} {2020})}\BibitemShut {NoStop}%
\bibitem [{\citenamefont {Uhrin}\ \emph {et~al.}(2024)\citenamefont {Uhrin}, \citenamefont {Zadoks}, \citenamefont {Binci}, \citenamefont {Marzari},\ and\ \citenamefont {Timrov}}]{Uhrin:2024}%
  \BibitemOpen
  \bibfield  {author} {\bibinfo {author} {\bibfnamefont {M.}~\bibnamefont {Uhrin}}, \bibinfo {author} {\bibfnamefont {A.}~\bibnamefont {Zadoks}}, \bibinfo {author} {\bibfnamefont {L.}~\bibnamefont {Binci}}, \bibinfo {author} {\bibfnamefont {N.}~\bibnamefont {Marzari}},\ and\ \bibinfo {author} {\bibfnamefont {I.}~\bibnamefont {Timrov}},\ }\bibfield  {title} {\bibinfo {title} {{Machine learning Hubbard parameters with equivariant neural networks}},\ }\href@noop {} {\bibfield  {journal} {\bibinfo  {journal} {arXiv:2406.02457}\ } (\bibinfo {year} {2024})}\BibitemShut {NoStop}%
\bibitem [{\citenamefont {Muraliganth}\ and\ \citenamefont {Manthiram}(2010)}]{Muraliganth:2010}%
  \BibitemOpen
  \bibfield  {author} {\bibinfo {author} {\bibfnamefont {T.}~\bibnamefont {Muraliganth}}\ and\ \bibinfo {author} {\bibfnamefont {A.}~\bibnamefont {Manthiram}},\ }\bibfield  {title} {\bibinfo {title} {{Understanding the Shifts in the Redox Potentials of Olivines LiM$_{1-y}$M$_y$PO$_4$ (M = Fe, Mn, Co, and Mg) Solid Solution Cathodes}},\ }\href@noop {} {\bibfield  {journal} {\bibinfo  {journal} {J. Phys. Chem. C}\ }\textbf {\bibinfo {volume} {114}},\ \bibinfo {pages} {15530} (\bibinfo {year} {2010})}\BibitemShut {NoStop}%
\bibitem [{\citenamefont {Kobayashi}\ \emph {et~al.}(2009)\citenamefont {Kobayashi}, \citenamefont {Yamada}, \citenamefont {ichi Nishimura}, \citenamefont {Kanno}, \citenamefont {Kobayashi}, \citenamefont {Seki}, \citenamefont {Ohno},\ and\ \citenamefont {Miyashiro}}]{KOBAYASHI2009}%
  \BibitemOpen
  \bibfield  {author} {\bibinfo {author} {\bibfnamefont {G.}~\bibnamefont {Kobayashi}}, \bibinfo {author} {\bibfnamefont {A.}~\bibnamefont {Yamada}}, \bibinfo {author} {\bibfnamefont {S.}~\bibnamefont {ichi Nishimura}}, \bibinfo {author} {\bibfnamefont {R.}~\bibnamefont {Kanno}}, \bibinfo {author} {\bibfnamefont {Y.}~\bibnamefont {Kobayashi}}, \bibinfo {author} {\bibfnamefont {S.}~\bibnamefont {Seki}}, \bibinfo {author} {\bibfnamefont {Y.}~\bibnamefont {Ohno}},\ and\ \bibinfo {author} {\bibfnamefont {H.}~\bibnamefont {Miyashiro}},\ }\bibfield  {title} {\bibinfo {title} {Shift of redox potential and kinetics in lix(mnyfe1-y)po4},\ }\href {https://doi.org/https://doi.org/10.1016/j.jpowsour.2008.07.085} {\bibfield  {journal} {\bibinfo  {journal} {J. Power Sources}\ }\textbf {\bibinfo {volume} {189}},\ \bibinfo {pages} {397} (\bibinfo {year} {2009})}\BibitemShut {NoStop}%
\bibitem [{\citenamefont {Bussi}\ \emph {et~al.}(2007)\citenamefont {Bussi}, \citenamefont {Donadio},\ and\ \citenamefont {Parrinello}}]{Bussi:2007}%
  \BibitemOpen
  \bibfield  {author} {\bibinfo {author} {\bibfnamefont {G.}~\bibnamefont {Bussi}}, \bibinfo {author} {\bibfnamefont {D.}~\bibnamefont {Donadio}},\ and\ \bibinfo {author} {\bibfnamefont {M.}~\bibnamefont {Parrinello}},\ }\bibfield  {title} {\bibinfo {title} {{Canonical sampling through velocity rescaling}},\ }\href {https://doi.org/10.1063/1.2408420} {\bibfield  {journal} {\bibinfo  {journal} {J. Chem. Phys.}\ }\textbf {\bibinfo {volume} {126}},\ \bibinfo {pages} {014101} (\bibinfo {year} {2007})}\BibitemShut {NoStop}%
\bibitem [{\citenamefont {Goodwin}\ \emph {et~al.}(2024)\citenamefont {Goodwin}, \citenamefont {Wenny}, \citenamefont {Yang}, \citenamefont {Cepellotti}, \citenamefont {Ding}, \citenamefont {Bystrom}, \citenamefont {Duschatko}, \citenamefont {Johansson}, \citenamefont {Sun}, \citenamefont {Batzner}, \citenamefont {Musaelian}, \citenamefont {Mason}, \citenamefont {Kozinsky},\ and\ \citenamefont {Molinari}}]{Goodwin:2024}%
  \BibitemOpen
  \bibfield  {author} {\bibinfo {author} {\bibfnamefont {Z.~A.~H.}\ \bibnamefont {Goodwin}}, \bibinfo {author} {\bibfnamefont {M.~B.}\ \bibnamefont {Wenny}}, \bibinfo {author} {\bibfnamefont {J.~H.}\ \bibnamefont {Yang}}, \bibinfo {author} {\bibfnamefont {A.}~\bibnamefont {Cepellotti}}, \bibinfo {author} {\bibfnamefont {J.}~\bibnamefont {Ding}}, \bibinfo {author} {\bibfnamefont {K.}~\bibnamefont {Bystrom}}, \bibinfo {author} {\bibfnamefont {B.~R.}\ \bibnamefont {Duschatko}}, \bibinfo {author} {\bibfnamefont {A.}~\bibnamefont {Johansson}}, \bibinfo {author} {\bibfnamefont {L.}~\bibnamefont {Sun}}, \bibinfo {author} {\bibfnamefont {S.}~\bibnamefont {Batzner}}, \bibinfo {author} {\bibfnamefont {A.}~\bibnamefont {Musaelian}}, \bibinfo {author} {\bibfnamefont {J.~A.}\ \bibnamefont {Mason}}, \bibinfo {author} {\bibfnamefont {B.}~\bibnamefont {Kozinsky}},\ and\ \bibinfo {author} {\bibfnamefont {N.}~\bibnamefont {Molinari}},\ }\bibfield  {title} {\bibinfo {title} {Transferability and accuracy of ionic liquid
  simulations with equivariant machine learning interatomic potentials},\ }\href {https://doi.org/10.1021/acs.jpclett.4c01942} {\bibfield  {journal} {\bibinfo  {journal} {J. Phys. Chem. Lett.}\ }\textbf {\bibinfo {volume} {15}},\ \bibinfo {pages} {7539} (\bibinfo {year} {2024})}\BibitemShut {NoStop}%
\bibitem [{\citenamefont {Perdew}\ \emph {et~al.}(1982)\citenamefont {Perdew}, \citenamefont {Parr}, \citenamefont {Levy},\ and\ \citenamefont {Balduz}}]{perdew:1982}%
  \BibitemOpen
  \bibfield  {author} {\bibinfo {author} {\bibfnamefont {J.~P.}\ \bibnamefont {Perdew}}, \bibinfo {author} {\bibfnamefont {R.~G.}\ \bibnamefont {Parr}}, \bibinfo {author} {\bibfnamefont {M.}~\bibnamefont {Levy}},\ and\ \bibinfo {author} {\bibfnamefont {J.~L.}\ \bibnamefont {Balduz}},\ }\bibfield  {title} {\bibinfo {title} {Density-{{Functional Theory}} for {{Fractional Particle Number}}: {{Derivative Discontinuities}} of the {{Energy}}},\ }\href {https://doi.org/10.1103/PhysRevLett.49.1691} {\bibfield  {journal} {\bibinfo  {journal} {Phys. Rev. Lett.}\ }\textbf {\bibinfo {volume} {49}},\ \bibinfo {pages} {1691} (\bibinfo {year} {1982})}\BibitemShut {NoStop}%
\bibitem [{\citenamefont {Cococcioni}(2002)}]{cococcioni:2002}%
  \BibitemOpen
  \bibfield  {author} {\bibinfo {author} {\bibfnamefont {M.}~\bibnamefont {Cococcioni}},\ }\emph {\bibinfo {title} {A {{LDA}}+{{U}} Study of Selected Iron Compounds}},\ \href@noop {} {Ph.D. thesis},\ \bibinfo  {school} {Scuola Internazionale Superiore di Studi Avanzati}, \bibinfo {address} {Trieste, ITA} (\bibinfo {year} {2002})\BibitemShut {NoStop}%
\bibitem [{\citenamefont {Cococcioni}\ and\ \citenamefont {de~Gironcoli}(2005)}]{Cococcioni:2005}%
  \BibitemOpen
  \bibfield  {author} {\bibinfo {author} {\bibfnamefont {M.}~\bibnamefont {Cococcioni}}\ and\ \bibinfo {author} {\bibfnamefont {S.}~\bibnamefont {de~Gironcoli}},\ }\bibfield  {title} {\bibinfo {title} {{Linear response approach to the calculation of the effective interaction parameters in the {LDA+U} method}},\ }\href@noop {} {\bibfield  {journal} {\bibinfo  {journal} {Phys. Rev. B}\ }\textbf {\bibinfo {volume} {71}},\ \bibinfo {pages} {035105} (\bibinfo {year} {2005})}\BibitemShut {NoStop}%
\bibitem [{\citenamefont {Kulik}\ \emph {et~al.}(2006)\citenamefont {Kulik}, \citenamefont {Cococcioni}, \citenamefont {Scherlis},\ and\ \citenamefont {Marzari}}]{Kulik:2006}%
  \BibitemOpen
  \bibfield  {author} {\bibinfo {author} {\bibfnamefont {H.}~\bibnamefont {Kulik}}, \bibinfo {author} {\bibfnamefont {M.}~\bibnamefont {Cococcioni}}, \bibinfo {author} {\bibfnamefont {D.}~\bibnamefont {Scherlis}},\ and\ \bibinfo {author} {\bibfnamefont {N.}~\bibnamefont {Marzari}},\ }\bibfield  {title} {\bibinfo {title} {{Density Functional Theory in Transition-Metal Chemistry: A Self-Consistent Hubbard U Approach}},\ }\href@noop {} {\bibfield  {journal} {\bibinfo  {journal} {Phys. Rev. Lett.}\ }\textbf {\bibinfo {volume} {97}},\ \bibinfo {pages} {103001} (\bibinfo {year} {2006})}\BibitemShut {NoStop}%
\bibitem [{\citenamefont {Mori-S\'anchez}\ \emph {et~al.}(2006)\citenamefont {Mori-S\'anchez}, \citenamefont {Cohen},\ and\ \citenamefont {Yang}}]{MoriSanchez:2006}%
  \BibitemOpen
  \bibfield  {author} {\bibinfo {author} {\bibfnamefont {P.}~\bibnamefont {Mori-S\'anchez}}, \bibinfo {author} {\bibfnamefont {A.}~\bibnamefont {Cohen}},\ and\ \bibinfo {author} {\bibfnamefont {W.}~\bibnamefont {Yang}},\ }\bibfield  {title} {\bibinfo {title} {{Many-electron self-interaction error in approximate density functionals}},\ }\href@noop {} {\bibfield  {journal} {\bibinfo  {journal} {J. Chem. Phys.}\ }\textbf {\bibinfo {volume} {125}},\ \bibinfo {pages} {201102} (\bibinfo {year} {2006})}\BibitemShut {NoStop}%
\bibitem [{\citenamefont {{Mori-S{\'a}nchez}}\ \emph {et~al.}(2009)\citenamefont {{Mori-S{\'a}nchez}}, \citenamefont {Cohen},\ and\ \citenamefont {Yang}}]{morisanchez:2009}%
  \BibitemOpen
  \bibfield  {author} {\bibinfo {author} {\bibfnamefont {P.}~\bibnamefont {{Mori-S{\'a}nchez}}}, \bibinfo {author} {\bibfnamefont {A.~J.}\ \bibnamefont {Cohen}},\ and\ \bibinfo {author} {\bibfnamefont {W.}~\bibnamefont {Yang}},\ }\bibfield  {title} {\bibinfo {title} {Discontinuous {{Nature}} of the {{Exchange-Correlation Functional}} in {{Strongly Correlated Systems}}},\ }\href {https://doi.org/10.1103/PhysRevLett.102.066403} {\bibfield  {journal} {\bibinfo  {journal} {Phys. Rev. Lett.}\ }\textbf {\bibinfo {volume} {102}},\ \bibinfo {pages} {066403} (\bibinfo {year} {2009})}\BibitemShut {NoStop}%
\bibitem [{\citenamefont {Zhao}\ \emph {et~al.}(2016)\citenamefont {Zhao}, \citenamefont {Ioannidis},\ and\ \citenamefont {Kulik}}]{zhao:2016}%
  \BibitemOpen
  \bibfield  {author} {\bibinfo {author} {\bibfnamefont {Q.}~\bibnamefont {Zhao}}, \bibinfo {author} {\bibfnamefont {E.~I.}\ \bibnamefont {Ioannidis}},\ and\ \bibinfo {author} {\bibfnamefont {H.~J.}\ \bibnamefont {Kulik}},\ }\bibfield  {title} {\bibinfo {title} {Global and local curvature in density functional theory},\ }\href {https://doi.org/10.1063/1.4959882} {\bibfield  {journal} {\bibinfo  {journal} {The Journal of Chemical Physics}\ }\textbf {\bibinfo {volume} {145}},\ \bibinfo {pages} {054109} (\bibinfo {year} {2016})}\BibitemShut {NoStop}%
\bibitem [{\citenamefont {Timrov}\ \emph {et~al.}(2020{\natexlab{a}})\citenamefont {Timrov}, \citenamefont {Aquilante}, \citenamefont {Binci}, \citenamefont {Cococcioni},\ and\ \citenamefont {Marzari}}]{Timrov:2020b}%
  \BibitemOpen
  \bibfield  {author} {\bibinfo {author} {\bibfnamefont {I.}~\bibnamefont {Timrov}}, \bibinfo {author} {\bibfnamefont {F.}~\bibnamefont {Aquilante}}, \bibinfo {author} {\bibfnamefont {L.}~\bibnamefont {Binci}}, \bibinfo {author} {\bibfnamefont {M.}~\bibnamefont {Cococcioni}},\ and\ \bibinfo {author} {\bibfnamefont {N.}~\bibnamefont {Marzari}},\ }\bibfield  {title} {\bibinfo {title} {{Pulay forces in density-functional theory with extended Hubbard functionals: from nonorthogonalized to orthogonalized manifolds}},\ }\href@noop {} {\bibfield  {journal} {\bibinfo  {journal} {Phys. Rev. B}\ }\textbf {\bibinfo {volume} {102}},\ \bibinfo {pages} {235159} (\bibinfo {year} {2020}{\natexlab{a}})},\ \bibinfo {note} {\textit{ibid.} \textbf{105}, 199901 (2022).}\BibitemShut {Stop}%
\bibitem [{\citenamefont {Shishkin}\ and\ \citenamefont {Sato}(2016)}]{Shishkin:2016}%
  \BibitemOpen
  \bibfield  {author} {\bibinfo {author} {\bibfnamefont {M.}~\bibnamefont {Shishkin}}\ and\ \bibinfo {author} {\bibfnamefont {H.}~\bibnamefont {Sato}},\ }\bibfield  {title} {\bibinfo {title} {{Self-consistent parametrization of DFT+$U$ framework using linear response approach: Application to evaluation of redox potentials of battery cathodes}},\ }\href@noop {} {\bibfield  {journal} {\bibinfo  {journal} {Phys. Rev. B}\ }\textbf {\bibinfo {volume} {93}},\ \bibinfo {pages} {085135} (\bibinfo {year} {2016})}\BibitemShut {NoStop}%
\bibitem [{\citenamefont {Timrov}\ \emph {et~al.}(2021)\citenamefont {Timrov}, \citenamefont {Marzari},\ and\ \citenamefont {Cococcioni}}]{Timrov:2021}%
  \BibitemOpen
  \bibfield  {author} {\bibinfo {author} {\bibfnamefont {I.}~\bibnamefont {Timrov}}, \bibinfo {author} {\bibfnamefont {N.}~\bibnamefont {Marzari}},\ and\ \bibinfo {author} {\bibfnamefont {M.}~\bibnamefont {Cococcioni}},\ }\bibfield  {title} {\bibinfo {title} {{Self-consistent Hubbard parameters from density-functional perturbation theory in the ultrasoft and projector-augmented wave formulations}},\ }\href@noop {} {\bibfield  {journal} {\bibinfo  {journal} {Phys. Rev. B}\ }\textbf {\bibinfo {volume} {103}},\ \bibinfo {pages} {045141} (\bibinfo {year} {2021})}\BibitemShut {NoStop}%
\bibitem [{\citenamefont {Timrov}\ \emph {et~al.}(2018)\citenamefont {Timrov}, \citenamefont {Marzari},\ and\ \citenamefont {Cococcioni}}]{Timrov:2018}%
  \BibitemOpen
  \bibfield  {author} {\bibinfo {author} {\bibfnamefont {I.}~\bibnamefont {Timrov}}, \bibinfo {author} {\bibfnamefont {N.}~\bibnamefont {Marzari}},\ and\ \bibinfo {author} {\bibfnamefont {M.}~\bibnamefont {Cococcioni}},\ }\bibfield  {title} {\bibinfo {title} {Hubbard parameters from density-functional perturbation theory},\ }\href@noop {} {\bibfield  {journal} {\bibinfo  {journal} {Phys. Rev. B}\ }\textbf {\bibinfo {volume} {98}},\ \bibinfo {pages} {085127} (\bibinfo {year} {2018})}\BibitemShut {NoStop}%
\bibitem [{\citenamefont {Hsu}\ \emph {et~al.}(2009)\citenamefont {Hsu}, \citenamefont {Umemoto}, \citenamefont {Cococcioni},\ and\ \citenamefont {Wentzcovitch}}]{Hsu:2009}%
  \BibitemOpen
  \bibfield  {author} {\bibinfo {author} {\bibfnamefont {H.}~\bibnamefont {Hsu}}, \bibinfo {author} {\bibfnamefont {K.}~\bibnamefont {Umemoto}}, \bibinfo {author} {\bibfnamefont {M.}~\bibnamefont {Cococcioni}},\ and\ \bibinfo {author} {\bibfnamefont {R.}~\bibnamefont {Wentzcovitch}},\ }\bibfield  {title} {\bibinfo {title} {{First-principles study for low-spin LaCoO$_3$ with a structurally consistent Hubbard $U$}},\ }\href@noop {} {\bibfield  {journal} {\bibinfo  {journal} {Phys. Rev. B}\ }\textbf {\bibinfo {volume} {79}},\ \bibinfo {pages} {125124} (\bibinfo {year} {2009})}\BibitemShut {NoStop}%
\bibitem [{\citenamefont {Timrov}\ \emph {et~al.}(2020{\natexlab{b}})\citenamefont {Timrov}, \citenamefont {Aquilante}, \citenamefont {Binci}, \citenamefont {Cococcioni},\ and\ \citenamefont {Marzari}}]{Timrov2020}%
  \BibitemOpen
  \bibfield  {author} {\bibinfo {author} {\bibfnamefont {I.}~\bibnamefont {Timrov}}, \bibinfo {author} {\bibfnamefont {F.}~\bibnamefont {Aquilante}}, \bibinfo {author} {\bibfnamefont {L.}~\bibnamefont {Binci}}, \bibinfo {author} {\bibfnamefont {M.}~\bibnamefont {Cococcioni}},\ and\ \bibinfo {author} {\bibfnamefont {N.}~\bibnamefont {Marzari}},\ }\bibfield  {title} {\bibinfo {title} {Pulay forces in density-functional theory with extended {Hubbard} functionals: From nonorthogonalized to orthogonalized manifolds},\ }\href {https://doi.org/10.1103/physrevb.102.235159} {\bibfield  {journal} {\bibinfo  {journal} {Physical Review B}\ }\textbf {\bibinfo {volume} {102}},\ \bibinfo {pages} {235159} (\bibinfo {year} {2020}{\natexlab{b}})}\BibitemShut {NoStop}%
\bibitem [{\citenamefont {Giannozzi}\ \emph {et~al.}(2009)\citenamefont {Giannozzi}, \citenamefont {Baroni}, \citenamefont {Bonini}, \citenamefont {Calandra}, \citenamefont {Car}, \citenamefont {Cavazzoni}, \citenamefont {Ceresoli}, \citenamefont {Chiarotti}, \citenamefont {Cococcioni}, \citenamefont {Dabo}, \citenamefont {Dal~Corso}, \citenamefont {De~Gironcoli}, \citenamefont {Fabris}, \citenamefont {Fratesi}, \citenamefont {Gebauer}, \citenamefont {Gerstmann}, \citenamefont {Gougoussis}, \citenamefont {Kokalj}, \citenamefont {Lazzeri}, \citenamefont {Martin-Samos}, \citenamefont {Marzari}, \citenamefont {Mauri}, \citenamefont {Mazzarello}, \citenamefont {Paolini}, \citenamefont {Pasquarello}, \citenamefont {Paulatto}, \citenamefont {Sbraccia}, \citenamefont {Scandolo}, \citenamefont {Sclauzero}, \citenamefont {Seitsonen}, \citenamefont {Smogunov}, \citenamefont {Umari},\ and\ \citenamefont {Wentzcovitch}}]{Giannozzi:2009}%
  \BibitemOpen
  \bibfield  {author} {\bibinfo {author} {\bibfnamefont {P.}~\bibnamefont {Giannozzi}}, \bibinfo {author} {\bibfnamefont {S.}~\bibnamefont {Baroni}}, \bibinfo {author} {\bibfnamefont {N.}~\bibnamefont {Bonini}}, \bibinfo {author} {\bibfnamefont {M.}~\bibnamefont {Calandra}}, \bibinfo {author} {\bibfnamefont {R.}~\bibnamefont {Car}}, \bibinfo {author} {\bibfnamefont {C.}~\bibnamefont {Cavazzoni}}, \bibinfo {author} {\bibfnamefont {D.}~\bibnamefont {Ceresoli}}, \bibinfo {author} {\bibfnamefont {G.}~\bibnamefont {Chiarotti}}, \bibinfo {author} {\bibfnamefont {M.}~\bibnamefont {Cococcioni}}, \bibinfo {author} {\bibfnamefont {I.}~\bibnamefont {Dabo}}, \bibinfo {author} {\bibfnamefont {A.}~\bibnamefont {Dal~Corso}}, \bibinfo {author} {\bibfnamefont {S.}~\bibnamefont {De~Gironcoli}}, \bibinfo {author} {\bibfnamefont {S.}~\bibnamefont {Fabris}}, \bibinfo {author} {\bibfnamefont {G.}~\bibnamefont {Fratesi}}, \bibinfo {author} {\bibfnamefont {R.}~\bibnamefont {Gebauer}}, \bibinfo {author} {\bibfnamefont
  {U.}~\bibnamefont {Gerstmann}}, \bibinfo {author} {\bibfnamefont {C.}~\bibnamefont {Gougoussis}}, \bibinfo {author} {\bibfnamefont {A.}~\bibnamefont {Kokalj}}, \bibinfo {author} {\bibfnamefont {M.}~\bibnamefont {Lazzeri}}, \bibinfo {author} {\bibfnamefont {L.}~\bibnamefont {Martin-Samos}}, \bibinfo {author} {\bibfnamefont {N.}~\bibnamefont {Marzari}}, \bibinfo {author} {\bibfnamefont {F.}~\bibnamefont {Mauri}}, \bibinfo {author} {\bibfnamefont {R.}~\bibnamefont {Mazzarello}}, \bibinfo {author} {\bibfnamefont {S.}~\bibnamefont {Paolini}}, \bibinfo {author} {\bibfnamefont {A.}~\bibnamefont {Pasquarello}}, \bibinfo {author} {\bibfnamefont {L.}~\bibnamefont {Paulatto}}, \bibinfo {author} {\bibfnamefont {C.}~\bibnamefont {Sbraccia}}, \bibinfo {author} {\bibfnamefont {S.}~\bibnamefont {Scandolo}}, \bibinfo {author} {\bibfnamefont {G.}~\bibnamefont {Sclauzero}}, \bibinfo {author} {\bibfnamefont {A.}~\bibnamefont {Seitsonen}}, \bibinfo {author} {\bibfnamefont {A.}~\bibnamefont {Smogunov}}, \bibinfo {author}
  {\bibfnamefont {P.}~\bibnamefont {Umari}},\ and\ \bibinfo {author} {\bibfnamefont {R.}~\bibnamefont {Wentzcovitch}},\ }\bibfield  {title} {\bibinfo {title} {{Q}uantum {ESPRESSO}: {A} modular and open-source software project for quantum simulations of materials},\ }\href@noop {} {\bibfield  {journal} {\bibinfo  {journal} {J. Phys.: Condens. Matter.}\ }\textbf {\bibinfo {volume} {21}},\ \bibinfo {pages} {395502} (\bibinfo {year} {2009})}\BibitemShut {NoStop}%
\bibitem [{\citenamefont {Giannozzi}\ \emph {et~al.}(2017)\citenamefont {Giannozzi}, \citenamefont {Andreussi}, \citenamefont {Brumme}, \citenamefont {Bunau}, \citenamefont {Buongiorno~Nardelli}, \citenamefont {Calandra}, \citenamefont {Car}, \citenamefont {Cavazzoni}, \citenamefont {Ceresoli}, \citenamefont {Cococcioni}, \citenamefont {Colonna}, \citenamefont {Carnimeo}, \citenamefont {Dal~Corso}, \citenamefont {de~Gironcoli}, \citenamefont {Delugas}, \citenamefont {Di{S}tasio~{J}r.}, \citenamefont {Ferretti}, \citenamefont {Floris}, \citenamefont {Fratesi}, \citenamefont {Fugallo}, \citenamefont {Gebauer}, \citenamefont {Gerstmann}, \citenamefont {Giustino}, \citenamefont {Gorni}, \citenamefont {Jia}, \citenamefont {Kawamura}, \citenamefont {Ko}, \citenamefont {Kokalj}, \citenamefont {K\"{u}\c{c}\"{u}kbenli}, \citenamefont {Lazzeri}, \citenamefont {Marsili}, \citenamefont {Marzari}, \citenamefont {Mauri}, \citenamefont {Nguyen}, \citenamefont {Nguyen}, \citenamefont {Otero-de-la {R}osa}, \citenamefont
  {Paulatto}, \citenamefont {Ponc\'e}, \citenamefont {Rocca}, \citenamefont {Sabatini}, \citenamefont {Santra}, \citenamefont {Schlipf}, \citenamefont {Seitsonen}, \citenamefont {Smogunov}, \citenamefont {Timrov}, \citenamefont {Thonhauser}, \citenamefont {Umari}, \citenamefont {Vast},\ and\ \citenamefont {Baroni}}]{Giannozzi:2017}%
  \BibitemOpen
  \bibfield  {author} {\bibinfo {author} {\bibfnamefont {P.}~\bibnamefont {Giannozzi}}, \bibinfo {author} {\bibfnamefont {O.}~\bibnamefont {Andreussi}}, \bibinfo {author} {\bibfnamefont {T.}~\bibnamefont {Brumme}}, \bibinfo {author} {\bibfnamefont {O.}~\bibnamefont {Bunau}}, \bibinfo {author} {\bibfnamefont {M.}~\bibnamefont {Buongiorno~Nardelli}}, \bibinfo {author} {\bibfnamefont {M.}~\bibnamefont {Calandra}}, \bibinfo {author} {\bibfnamefont {R.}~\bibnamefont {Car}}, \bibinfo {author} {\bibfnamefont {C.}~\bibnamefont {Cavazzoni}}, \bibinfo {author} {\bibfnamefont {D.}~\bibnamefont {Ceresoli}}, \bibinfo {author} {\bibfnamefont {M.}~\bibnamefont {Cococcioni}}, \bibinfo {author} {\bibfnamefont {N.}~\bibnamefont {Colonna}}, \bibinfo {author} {\bibfnamefont {I.}~\bibnamefont {Carnimeo}}, \bibinfo {author} {\bibfnamefont {A.}~\bibnamefont {Dal~Corso}}, \bibinfo {author} {\bibfnamefont {S.}~\bibnamefont {de~Gironcoli}}, \bibinfo {author} {\bibfnamefont {P.}~\bibnamefont {Delugas}}, \bibinfo {author} {\bibfnamefont
  {R.~A.}\ \bibnamefont {Di{S}tasio~{J}r.}}, \bibinfo {author} {\bibfnamefont {A.}~\bibnamefont {Ferretti}}, \bibinfo {author} {\bibfnamefont {A.}~\bibnamefont {Floris}}, \bibinfo {author} {\bibfnamefont {G.}~\bibnamefont {Fratesi}}, \bibinfo {author} {\bibfnamefont {G.}~\bibnamefont {Fugallo}}, \bibinfo {author} {\bibfnamefont {R.}~\bibnamefont {Gebauer}}, \bibinfo {author} {\bibfnamefont {U.}~\bibnamefont {Gerstmann}}, \bibinfo {author} {\bibfnamefont {F.}~\bibnamefont {Giustino}}, \bibinfo {author} {\bibfnamefont {T.}~\bibnamefont {Gorni}}, \bibinfo {author} {\bibfnamefont {J.}~\bibnamefont {Jia}}, \bibinfo {author} {\bibfnamefont {M.}~\bibnamefont {Kawamura}}, \bibinfo {author} {\bibfnamefont {H.-Y.}\ \bibnamefont {Ko}}, \bibinfo {author} {\bibfnamefont {A.}~\bibnamefont {Kokalj}}, \bibinfo {author} {\bibfnamefont {E.}~\bibnamefont {K\"{u}\c{c}\"{u}kbenli}}, \bibinfo {author} {\bibfnamefont {M.}~\bibnamefont {Lazzeri}}, \bibinfo {author} {\bibfnamefont {M.}~\bibnamefont {Marsili}}, \bibinfo {author}
  {\bibfnamefont {N.}~\bibnamefont {Marzari}}, \bibinfo {author} {\bibfnamefont {F.}~\bibnamefont {Mauri}}, \bibinfo {author} {\bibfnamefont {N.~L.}\ \bibnamefont {Nguyen}}, \bibinfo {author} {\bibfnamefont {H.-V.}\ \bibnamefont {Nguyen}}, \bibinfo {author} {\bibfnamefont {A.}~\bibnamefont {Otero-de-la {R}osa}}, \bibinfo {author} {\bibfnamefont {L.}~\bibnamefont {Paulatto}}, \bibinfo {author} {\bibfnamefont {S.}~\bibnamefont {Ponc\'e}}, \bibinfo {author} {\bibfnamefont {D.}~\bibnamefont {Rocca}}, \bibinfo {author} {\bibfnamefont {R.}~\bibnamefont {Sabatini}}, \bibinfo {author} {\bibfnamefont {B.}~\bibnamefont {Santra}}, \bibinfo {author} {\bibfnamefont {M.}~\bibnamefont {Schlipf}}, \bibinfo {author} {\bibfnamefont {A.}~\bibnamefont {Seitsonen}}, \bibinfo {author} {\bibfnamefont {A.}~\bibnamefont {Smogunov}}, \bibinfo {author} {\bibfnamefont {I.}~\bibnamefont {Timrov}}, \bibinfo {author} {\bibfnamefont {T.}~\bibnamefont {Thonhauser}}, \bibinfo {author} {\bibfnamefont {P.}~\bibnamefont {Umari}}, \bibinfo
  {author} {\bibfnamefont {N.}~\bibnamefont {Vast}},\ and\ \bibinfo {author} {\bibfnamefont {S.}~\bibnamefont {Baroni}},\ }\bibfield  {title} {\bibinfo {title} {{A}dvanced capabilities for materials modelling with {Q}uantum {ESPRESSO}},\ }\href {https://doi.org/10.1088/1361-648X/aa8f79} {\bibfield  {journal} {\bibinfo  {journal} {J. Phys.: Condens. Matter.}\ }\textbf {\bibinfo {volume} {29}},\ \bibinfo {pages} {465901} (\bibinfo {year} {2017})}\BibitemShut {NoStop}%
\bibitem [{\citenamefont {Perdew}\ \emph {et~al.}(2008)\citenamefont {Perdew}, \citenamefont {Ruzsinszky}, \citenamefont {Csonka}, \citenamefont {Vydrov}, \citenamefont {Scuseria}, \citenamefont {Constantin}, \citenamefont {Zhou},\ and\ \citenamefont {Burke}}]{Perdew:2008}%
  \BibitemOpen
  \bibfield  {author} {\bibinfo {author} {\bibfnamefont {J.}~\bibnamefont {Perdew}}, \bibinfo {author} {\bibfnamefont {A.}~\bibnamefont {Ruzsinszky}}, \bibinfo {author} {\bibfnamefont {G.}~\bibnamefont {Csonka}}, \bibinfo {author} {\bibfnamefont {O.}~\bibnamefont {Vydrov}}, \bibinfo {author} {\bibfnamefont {G.}~\bibnamefont {Scuseria}}, \bibinfo {author} {\bibfnamefont {L.}~\bibnamefont {Constantin}}, \bibinfo {author} {\bibfnamefont {X.}~\bibnamefont {Zhou}},\ and\ \bibinfo {author} {\bibfnamefont {K.}~\bibnamefont {Burke}},\ }\href@noop {} {\bibfield  {journal} {\bibinfo  {journal} {Phys. Rev. Lett.}\ }\textbf {\bibinfo {volume} {100}},\ \bibinfo {pages} {136406} (\bibinfo {year} {2008})}\BibitemShut {NoStop}%
\bibitem [{Mat()}]{MaterialsCloud}%
  \BibitemOpen
  \href@noop {} {}\bibinfo {note} {{The SSSP library of the Materials Cloud: \url{https://www.materialscloud.org/discover/sssp/table/precision}}}\BibitemShut {NoStop}%
\bibitem [{\citenamefont {Garrity}\ \emph {et~al.}(2014)\citenamefont {Garrity}, \citenamefont {Bennett}, \citenamefont {Rabe},\ and\ \citenamefont {Vanderbilt}}]{Garrity:2014}%
  \BibitemOpen
  \bibfield  {author} {\bibinfo {author} {\bibfnamefont {K.}~\bibnamefont {Garrity}}, \bibinfo {author} {\bibfnamefont {J.}~\bibnamefont {Bennett}}, \bibinfo {author} {\bibfnamefont {K.}~\bibnamefont {Rabe}},\ and\ \bibinfo {author} {\bibfnamefont {D.}~\bibnamefont {Vanderbilt}},\ }\href@noop {} {\bibfield  {journal} {\bibinfo  {journal} {Comput. Mater. Sci.}\ }\textbf {\bibinfo {volume} {81}},\ \bibinfo {pages} {446} (\bibinfo {year} {2014})}\BibitemShut {NoStop}%
\bibitem [{\citenamefont {{Dal Corso}}(2014)}]{DalCorso:2014}%
  \BibitemOpen
  \bibfield  {author} {\bibinfo {author} {\bibfnamefont {A.}~\bibnamefont {{Dal Corso}}},\ }\bibfield  {title} {\bibinfo {title} {{Pseudopotentials periodic table: From H to Pu}},\ }\href@noop {} {\bibfield  {journal} {\bibinfo  {journal} {Comput. Mater. Sci.}\ }\textbf {\bibinfo {volume} {95}},\ \bibinfo {pages} {337} (\bibinfo {year} {2014})}\BibitemShut {NoStop}%
\bibitem [{\citenamefont {Timrov}\ \emph {et~al.}(2022{\natexlab{b}})\citenamefont {Timrov}, \citenamefont {Marzari},\ and\ \citenamefont {Cococcioni}}]{Timrov:2022}%
  \BibitemOpen
  \bibfield  {author} {\bibinfo {author} {\bibfnamefont {I.}~\bibnamefont {Timrov}}, \bibinfo {author} {\bibfnamefont {N.}~\bibnamefont {Marzari}},\ and\ \bibinfo {author} {\bibfnamefont {M.}~\bibnamefont {Cococcioni}},\ }\bibfield  {title} {\bibinfo {title} {{\texttt{HP} -- A code for the calculation of Hubbard parameters using density-functional perturbation theory}},\ }\href@noop {} {\bibfield  {journal} {\bibinfo  {journal} {Comput. Phys. Commun.}\ }\textbf {\bibinfo {volume} {279}},\ \bibinfo {pages} {108455} (\bibinfo {year} {2022}{\natexlab{b}})}\BibitemShut {NoStop}%
\bibitem [{\citenamefont {Fletcher}(1987)}]{Fletcher:1987}%
  \BibitemOpen
  \bibfield  {author} {\bibinfo {author} {\bibfnamefont {R.}~\bibnamefont {Fletcher}},\ }\href@noop {} {\emph {\bibinfo {title} {{Practical Methods of Optimization}}}},\ \bibinfo {edition} {2nd}\ ed.\ (\bibinfo  {publisher} {Wiley},\ \bibinfo {address} {Chichester},\ \bibinfo {year} {1987})\BibitemShut {NoStop}%
\bibitem [{\citenamefont {Malica}\ and\ \citenamefont {Marzari}(2024)}]{malica2024_materialscloud}%
  \BibitemOpen
  \bibfield  {author} {\bibinfo {author} {\bibfnamefont {C.}~\bibnamefont {Malica}}\ and\ \bibinfo {author} {\bibfnamefont {N.}~\bibnamefont {Marzari}},\ }\href {https://doi.org/10.24435/materialscloud:w7-k1} {\bibinfo {title} {Teaching oxidation states to neural networks}},\ \bibinfo {howpublished} {\url{https://doi.org/10.24435/materialscloud:w7-k1}} (\bibinfo {year} {2024}),\ \bibinfo {note} {materials Cloud Archive 2024.189}\BibitemShut {NoStop}%
\end{thebibliography}%


\begin{thebibliography}{1}
\urlstyle{rm}
\expandafter\ifx\csname url\endcsname\relax
  \def\url#1{\texttt{#1}}\fi
\expandafter\ifx\csname urlprefix\endcsname\relax\def\urlprefix{URL }\fi
\expandafter\ifx\csname doiprefix\endcsname\relax\def\doiprefix{DOI: }\fi
\providecommand{\bibinfo}[2]{#2}
\providecommand{\eprint}[2][]{\url{#2}}

\bibitem{Timrov:2022b}
\bibinfo{author}{Timrov, I.}, \bibinfo{author}{Aquilante, F.}, \bibinfo{author}{Cococcioni, M.} \& \bibinfo{author}{Marzari, N.}
\newblock \bibinfo{journal}{\bibinfo{title}{Accurate electronic properties and intercalation voltages of olivine-type li-ion cathode materials from extended hubbard functionals}}.
\newblock {\emph{\JournalTitle{Phys. Rev. X Energy}}} \textbf{\bibinfo{volume}{1}}, \bibinfo{pages}{033003}, \doiprefix\url{10.1103/PRXEnergy.1.033003} (\bibinfo{year}{2022}).

\bibitem{Timrov:2021}
\bibinfo{author}{Timrov, I.}, \bibinfo{author}{Marzari, N.} \& \bibinfo{author}{Cococcioni, M.}
\newblock \bibinfo{journal}{\bibinfo{title}{{Self-consistent Hubbard parameters from density-functional perturbation theory in the ultrasoft and projector-augmented wave formulations}}}.
\newblock {\emph{\JournalTitle{Phys. Rev. B}}} \textbf{\bibinfo{volume}{103}}, \bibinfo{pages}{045141} (\bibinfo{year}{2021}).

\bibitem{Sit:2011}
\bibinfo{author}{Sit, P. H.-L.}, \bibinfo{author}{Car, R.}, \bibinfo{author}{Cohen, M.~H.} \& \bibinfo{author}{Selloni, A.}
\newblock \bibinfo{journal}{\bibinfo{title}{{Simple, Unambiguous Theoretical Approach to Oxidation State Determination via First-Principles Calculations}}}.
\newblock {\emph{\JournalTitle{Inorg. Chem.}}} \textbf{\bibinfo{volume}{50}}, \bibinfo{pages}{10259} (\bibinfo{year}{2011}).

\end{thebibliography}




\end{document}



\title{
    \large{
        Supplementary Information\\
        Teaching oxidation states to neural networks
    }
}

\author{Cristiano Malica}
\email{cmalica@uni-bremen.de}
\affiliation{U Bremen Excellence Chair, Bremen Center for Computational Materials Science, and MAPEX Center for Materials and Processes, University of Bremen, D-28359 Bremen, Germany}
% ...
\author{Nicola Marzari}
\affiliation{U Bremen Excellence Chair, Bremen Center for Computational Materials Science, and MAPEX Center for Materials and Processes, University of Bremen, D-28359 Bremen, Germany}
\affiliation{Laboratory for Materials Simulations, Paul Scherrer Institut, 5232 Villigen PSI, Switzerland}
\affiliation{Theory and Simulation of Materials (THEOS), and National Centre for Computational Design and Discovery of Novel Materials (MARVEL), \'Ecole Polytechnique F\'ed\'erale de Lausanne (EPFL), CH-1015 Lausanne, Switzerland}

\date{\today}
\maketitle

\clearpage

\clearpage

\section{Self-Consistent Hubbard U and V Parameters of lmpo}
In the following Table~\ref{table:HP} we report the whole set of self-consistent Hubbard parameters for the LMPO phospho-olivne system (see also Ref.~\cite{Timrov:2022b}).
The self-consistent Hubbard parameters are computed by alternating structural optimization to the linear response calculation of Hubbard U and V parameters by using density-functional perturbation theory (DFPT) up to reach a convergence of 0.1 eV on both parameters. The self-consistent procedure is explained in more detail in Refs.~\cite{Timrov:2021}.   
The self-consistent Hubbard parameters reported in Table~\ref{table:HP} are used in the DFT+U+V calculation to derive electronic occupations and oxidation states in Fig. 1 (b) and Voltages in Fig. 1 (d) as discussed in the main text. 

\begin{table}[h!]
\centering
\begin{tabular}{|c|c|c|c|c|c|}
\hline
$x$ & HP & Mn1 & Mn2 & Mn3 & Mn4 \\ \hline
0 & \(U\) & 6.26 & 6.26 & 6.26 & 6.26 \\ 
  & \(V\) & 0.54-1.07 & 0.54-1.07 & 0.54-1.07 & 0.54-1.07 \\ \hline
1/4 & \(U\) & 6.26 & 6.25 & 6.67 & 5.44 \\ 
    & \(V\) & 0.40-1.01 & 0.46-1.05 & 0.54-1.11 & 0.39-1.08 \\ \hline
1/2 & \(U\) & 6.42 & 4.95 & 6.41 & 4.94 \\ 
    & \(V\) & 0.34-1.01 & 0.38-0.96 & 0.34-1.01 & 0.38-0.96 \\ \hline
3/4 & \(U\) & 4.67 & 4.64 & 6.58 & 4.98 \\ 
    & \(V\) & 0.48-0.72 & 0.31-0.91 & 0.33-1.02 & 0.41-0.79 \\ \hline
1 & \(U\) & 4.56 & 4.56 & 4.56 & 4.56 \\ 
  & \(V\) & 0.42-0.78 & 0.42-0.78 & 0.42-0.78 & 0.42-0.78 \\ \hline
\end{tabular}
\caption{Self-consistent Hubbard parameters (HP) in eV computed using DFPT in the DFT+U+V framework for $3d$-states of Mn in Li$_x$MnPO$_4$ for x=0, 1/4, 1/2, 3/4, 1.}
\label{table:HP}
\end{table}

\section{Oxidation states and electronic occupations}

\begin{figure}[h!]
    \centering
    \includegraphics[width=0.95\textwidth]{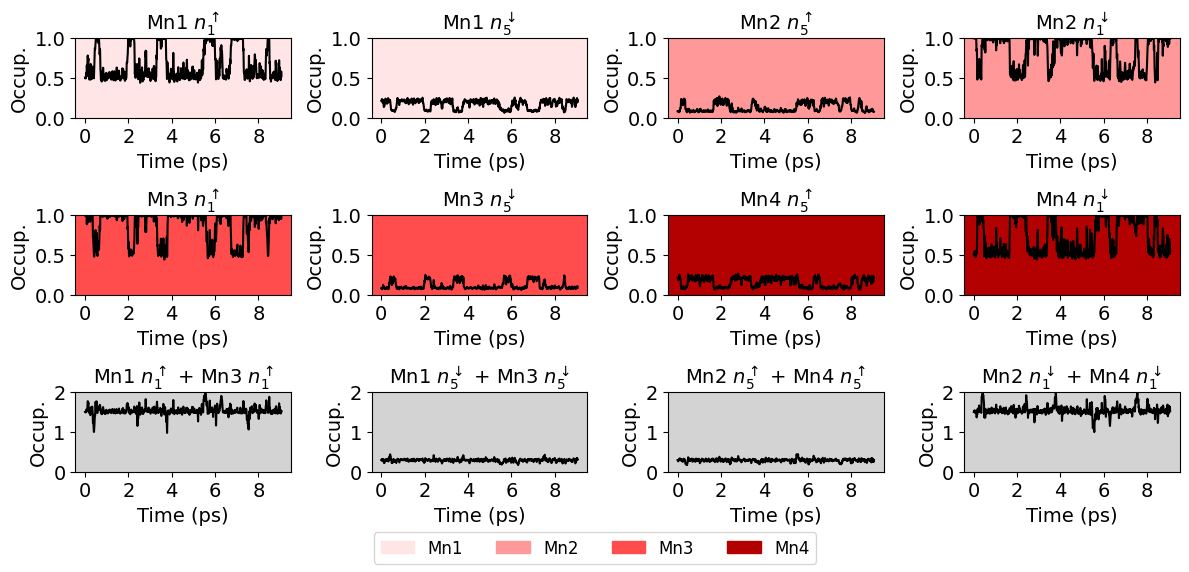}
    \caption{
        \textbf{Time evolution of the Mn-$3d$ electronic occupations for the four Mn atoms in the unit cell of L$_{x}$MnPO$_4$ (x=0.5) in DFT+U+V FPMD}. Different shades distinguish each atom, with grey-background plots indicating the sum of occupations from the plots above.
    }
    \label{fig:supp1}
\end{figure}

In the main text we have used L\"owdin occupations of Mn-$3d$ states (i.e., the sum of all electronic occupations) to monitor the variation of oxidation states (OSs) of Mn atoms during the time. This is, for example, shown in Figs. 2 (a-c) of the main text for Li$_x$MnPO$_4$ with $x=1/2$. Here, we analyze the behavior of the corresponding single electronic occupations over time. In passing, we recall that, within this Li concentration, there are two additional electrons that can localize in two different Mn atoms. 
In Fig.~\ref{fig:supp1} we report the corresponding eigenvalues of the atomic occupation matrix for the Mn-$3d$ states as a function of the time. In particular, we show the electronic occupations having visible variation over time for the four Mn atoms present in the unit cell. We observe that, the transitions in OSs is mainly determined by one of them, namely $n_1^{\uparrow}$ for Mn1 and Mn3, $n_1^{\downarrow}$ for Mn2 and Mn4 (where $\uparrow$ and $\downarrow$ indicate the spin up and spin down channels, respectively). These occupations undergo into net and sharp transitions of average amplitude $\sim 0.5$. Smaller variations are also visible in $n_5^{\downarrow}$ for Mn1 and Mn3 and in $n_5^{\uparrow}$ for Mn2 and Mn4. The other electronic occupations, not shown in the plot, remain stationary over time with small thermal fluctuations. 
For example, let us consider the case of $n_1^{\uparrow}$ of Mn1 and Mn3 with parallel spins. They are jumping approximately between 0.5 and 1.0 indicating that ---by following the method of Ref.~\cite{Sit:2011}--- the corresponding Mn atoms are shifting their OSs between 3+ and 2+. The time evolution of their sum (underlying grey plot) fluctuates in average around 1.5, meaning that when one occupation is 1 (OS 2+) the other is 0.5 (OS 3+). Sporadic configurations show a sum of 2 or 1 meaning that both electrons are localized in Mn1 and Mn3 or they are localized in Mn2 and Mn4, respectively. Similar considerations are valid for $n_1^{\downarrow}$ of Mn2 and Mn3. Hence, the adiabatic electron movement is preferred between atoms having the same spin. The sum of other occupations is also constant, meaning that there is no dispersion of charge in the environment.  

\begin{figure}[h!]
    \centering
    \includegraphics[width=0.5\textwidth]{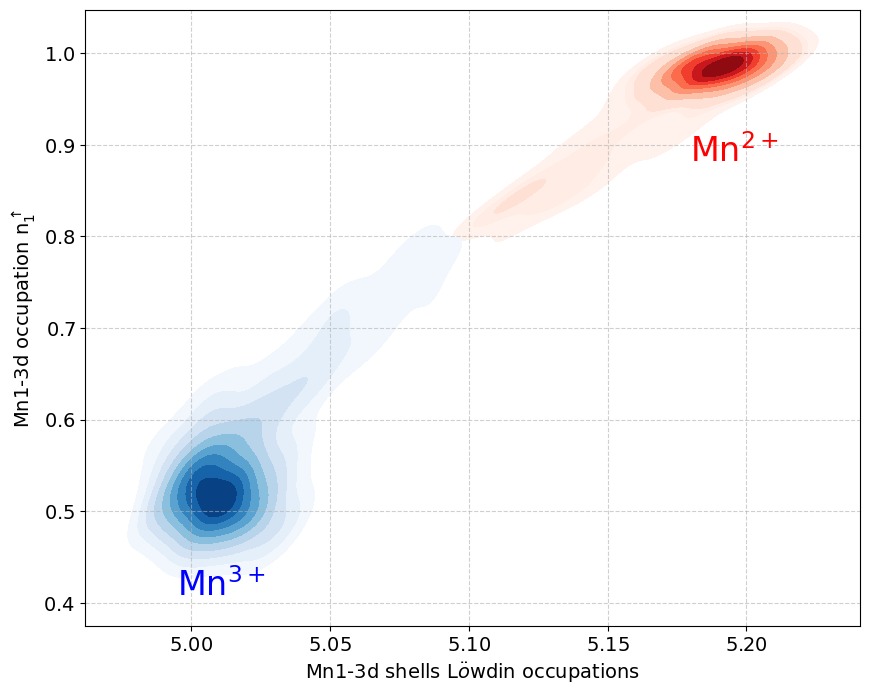}
    \caption{
        \textbf{Correlation between $n_1^{\uparrow}$ and L\"owdin occupations in the DFT+U+V FPMD of Li$_x$MnPO$_4$ with x=1/2 at T=900 K}. The intensity of the color increases with the number of configurations.
    }
    \label{fig:supp2}
\end{figure}

Finally, the correlation of $n_1^{\uparrow}$ and relative L\"owdin occupations of Mn1 is presented in Fig.~\ref{fig:supp2} for the whole set of configurations of the trajectory at T=900 K. The coefficient of linear correlation is 0.97 meaning, once again, that the total variation in the charge---measured by the L\"owdin occupations---is meanly related to the variation of $n_1^{\uparrow}$. The configurations with intermediate occupations are fewer, the ones with "clear" oxidation states are preferred. 

\section{Workflow for Selecting Data For Training}

\begin{figure}[b]
    \centering
    \includegraphics[width=0.25\textwidth]{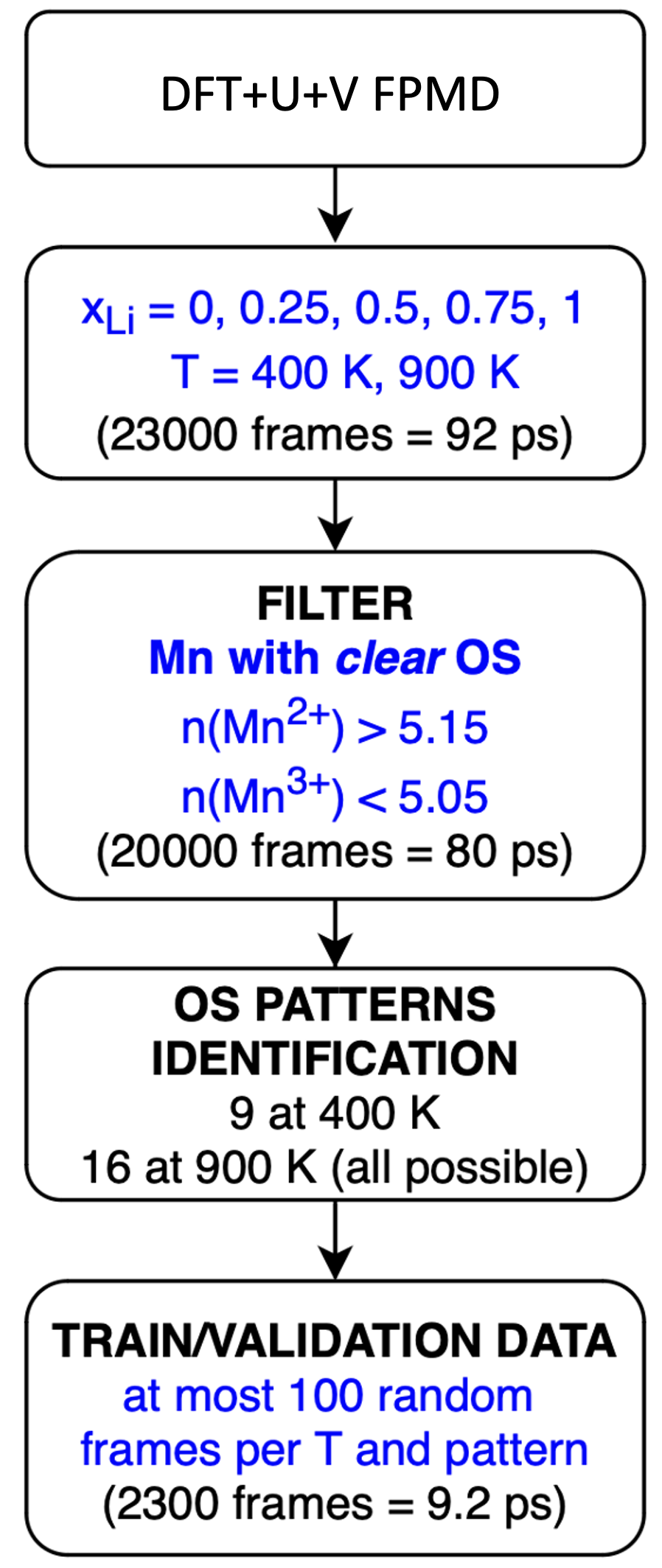}
    \caption{Workflow for the determination of the training/validation sets of atomic configurations for the generation of the machine learning interatomic potential. The FPMD is performed for all x$_{Li}$ concentration of Li at two different temperature, T=400 K and T=900 K, in the NVT ensemble.}
    \label{fig:wf_nn}
\end{figure}

In Fig.~\ref{fig:wf_nn} we show a schematic representation of the procedure for selecting training and validation data described in the main text of the manuscript. 

\clearpage

%

\clearpage
\section{Supplementary References}
\bibliography{si}
